\documentclass[usenatbib]{mn2e}
\usepackage{epsfig,rotating,natbib,amssymb}

\def\bv{\mathbf{v}}
\def\bA{\mathbf{A}}
\def\bB{\mathbf{B}}
\def\br{\mathbf{r}}

\def\curl{\nabla\times}

\def\smb{\sum_b m_b}
\def\smc{\sum_c m_c}
\def\gwab{\nabla_a W_{ab}}
\def\gawab{\nabla_a W_{ab}}
\def\divB{\nabla\cdot\mathbf{B}}
\def\runit{\hat{\bf r}}
\newcommand{\pder}[2]{\ensuremath{\frac{\partial #1}{\partial #2}}}

\title[Smoothed Particle Magnetohydrodynamics III]{Smoothed Particle
Magnetohydrodynamics \\ III. Multidimensional tests and the $\divB = 0$ constraint}

\author[Price]{D.J. Price$^{1}$, J.J. Monaghan$^{2}$\\
$^1$School of Physics, University of Exeter, Stocker Rd, Exeter EX4 4QL\\
$^2$School of Mathematical Sciences, Monash University, Clayton 3800, Australia\\
}
\date{Submitted: 9th June 2005 Revised: 30th August 2005}
\pagerange{\pageref{firstpage}--\pageref{lastpage}} \pubyear{2004}

\begin{document}
\label{firstpage}
\bibliographystyle{mn2e}
\maketitle

\begin{abstract}
 In two previous papers \citep{pm04a,pm04b} (papers I,II) we have described an algorithm for solving the equations of
Magnetohydrodynamics (MHD) using the Smoothed Particle Hydrodynamics (SPH)
method. The algorithm uses dissipative terms in order to capture shocks and has been tested on a wide range of one dimensional
problems in both adiabatic and isothermal MHD. In this paper we investigate multidimensional aspects of the
algorithm, refining many of the aspects considered in papers I and II and paying
particular attention to the code's ability to maintain the $\divB = 0$
constraint associated with the magnetic field. In particular we implement a hyperbolic divergence cleaning method recently
proposed by \citet{dea02} in combination with the consistent formulation of the
MHD equations in the presence of non-zero magnetic divergence derived in papers I and
II. Various projection methods for maintaining the divergence-free condition are also
examined. Finally the algorithm is tested against a wide range of multidimensional problems used to test
recent grid-based MHD codes. A particular finding of these tests is that in SPMHD the magnitude of the divergence error is dependent on the number of neighbours used to calculate a particle's properties and only weakly dependent on the total number of particles. 
Whilst many improvements could still be made to the algorithm, our results suggest that the method is ripe for application to problems of current theoretical interest, such as that of star formation.
\end{abstract}


%----------------------------------------------------------------------------------------------------------------
\section{Introduction}
 Magnetic fields play an important, in some cases crucial, role in many areas
of astrophysics. Despite the relative simplicity and well-studied nature of the
equations which describe them, their effects are complicated and analytic
studies are difficult and limited in scope. It is for this reason that a large
theoretical effort over the past decade or so has been devoted to developing
accurate numerical algorithms for Magnetohydrodynamics (MHD) in an
astrophysical context. There are, however, severe technical challenges to be
overcome in the numerical solution of the MHD equations.

 Smoothed Particle Hydrodynamics (SPH, for a review see \citealt{monaghan92}) is
a fully Lagrangian particle method which solves the equations of fluid dynamics
on a system of moving interpolation points which follow the fluid motion. SPH is
an extremely versatile and robust numerical method and as a result has found
widespread use in Astrophysics. There have, however, been difficulties with previous
attempts to simulate magnetic fields within SPH, most prominently due to a
numerical instability found to occur when an exactly momentum conserving form of
the SPMHD equations was used.

 In two previous papers (\citealt{pm04a,pm04b}, hereafter papers I and II), we have
described an algorithm for SPMHD in detail. The discrete
equations are formulated from a variational principle (paper~II) which ensures
consistency with physical principles (such as conservation of momentum and
energy) and a consistent treatment of magnetic divergence terms, the effects of
which we will investigate in this paper. Artificial dissipation terms
appropriate for shock-type problems were formulated in paper~I. These terms are carefully
formulated to give a positive definite contribution to the entropy. The
algorithm has been tested on a wide range of standard one dimensional problems
used to test recent grid-based MHD codes and also on the one dimensional `Toy Stars' of \citet{mp04}. The algorithm has been shown to give robust and accurate results on these problems.

 In more than one spatial dimension errors associated with the non-zero divergence of
the magnetic field need to be taken into account in any numerical MHD scheme.
There are two distinct issues to be addressed. The first is the treatment of
terms proportional to $\divB$ in the MHD equations (in particular in the
formulation of the induction equation and the magnetic force). The second is
the maintenance of the $\divB = 0$ constraint. It should be noted that a solution to the latter
problem does not necessarily resolve the former, since maintaining $\divB = 0$ in
a particular numerical discretisation does not guarantee that it is zero in all
discretisations.

  With regards to the first issue, \citet{bb80} first noted that, when using a conservative formulation of the magnetic force, a supposed steady state could become polluted because of the small but non-zero component of magnetic force directed along
the field lines due to a non-zero $\divB$. This error can have serious consequences
even though the proportional error in the magnetic field is small. In SPMHD the force
parallel to the field in conservative formulations can have catastrophic
consequences, leading to numerical instability under some circumstances
\citep{pm85}. \citet{bb80} approached
this problem by preferring a non-conservative formulation of the momentum equation which guarantees
that the magnetic force is exactly perpendicular to the field. Such an approach has also been used
successfully in an SPMHD context by several authors \citep[e.g.][]{benz84,mwd95,bp96,cg01}, however
numerical simulations of shocks seem to require the exact conservation of momentum in order to
provide the correct jump conditions at shock fronts (which means, at the very least, the discrete
formulation should be based on continuum equations which conserve momentum exactly even with a
non-zero magnetic divergence). 

 This issue of neglect or inclusion of terms proportional to $\divB$ was discussed at some length in paper~I, where we followed both \citet{janhunen00} and \citet{dellar01} in formulating the MHD equations such that they form a consistent set in the presence of magnetic monopole terms, retaining both
the conservation of momentum and energy necessary for shocks but using a
`non-conservative' formulation of the induction equation. The SPMHD equation set used
in papers I was shown to form a consistent set with respect to the monopole terms in both discrete and continuum forms by deriving the SPMHD equations of motion and energy from a variational principle which uses the discrete formulations of the continuity and induction equations as constraints (see paper~II). The
implications of the formulation of the MHD equations in the propagation of divergence
errors is discussed further in \S\ref{sec:monopoles} and examined numerically in \S\ref{sec:Bxpeaktest}.

 Many approaches to the second issue (namely the maintenance of the $\divB=0$ constraint) are possible. Perhaps
the simplest in an MHD context is to explicitly evolve a vector potential ${\bf A}$, from which
the magnetic field is derived by taking the curl, guaranteeing that the
divergence is zero. The major disadvantage of this
approach is that the computation of the force terms involves second
derivatives of the evolved variable ({\bf A}), which in general can be
significantly less accurate. Furthermore evaluation of dissipative terms proportional to $\nabla^2 {\bf B}$ would require computation of the \emph{third} derivatives. Whilst it may be possible to use the vector potential in an SPMHD context without degrading the accuracy substantially, we do not pursue such an investigation in this paper (although it is our intention to do so elsewhere).

 \citet{bb80} proposed a simple projection scheme to `clean up' the
magnetic field at each timestep, an approach which is now commonly used in many grid-based MHD codes
\citep[e.g.][]{balsara98}. Similar schemes have been implemented in an SPH context for
the simulation of incompressible
flows \citep{cr99}. The disadvantage of this approach is that it involves
the solution of a Poisson equation which is computationally expensive. In self-gravitating SPMHD this approach holds some promise as the cost may be mitigated by utilising the treecode used in the calculation of the gravitational force. In this paper we
examine various projection methods based on this approach in \S\ref{sec:projection}.

 Another
approach used in grid-based MHD codes is the so-called `constrained transport'
method pioneered by \citet{eh88} in which differences of the magnetic field
across the grid cell are constructed in such a way as to maintain the divergence
free condition exactly. Such methods work very well, but is difficult to see how they can be made applicable to SPH because of the absence of a spatial grid (although perhaps some
divergence-free interpolation could be devised). A comparison between several
constrained-transport type schemes with the source term approach of
\citet{pea99} and the projection method has been recently presented by
\citet{toth00} for finite difference codes. Although not all of the schemes are applicable in an SPH
context, many of the numerical tests presented in this chapter are taken
from \citeauthor{toth00}'s paper.

 More recently \citet{dea02} have proposed a method for cleaning
the magnetic field which is significantly faster than the projection method. This
method involves explicitly adding a constraint propagation equation which is coupled to the
evolution equation for the magnetic field. This equation propagates the divergence error in
a hyperbolic (ie. wave-like) manner away from its source, allowing diffusion of the
error to proceed rapidly within the timestep condition. This method is easily
applied to the SPMHD algorithm and we provide details of the implementation in
\S\ref{sec:hyperbolic}. 

 The paper is organised as follows: In \S\ref{sec:continuum} and \S\ref{sec:spmhd} we
summarise the formulation of the continuum MHD equations and the corresponding SPMHD
form of these equations from papers I and II. In the course of the multidimensional
testing, several aspects of the algorithm have been changed or refined from that
presented in papers I and II. The first change is the method for removing the
tensile instability, which is therefore discussed in \S\ref{sec:instability}. The
implementation of the dissipative terms formulated in paper~I in order to
capture shocks (paper~I) is reviewed and modified accordingly in \S\ref{sec:mhdav}. In
\S\ref{sec:divb}  we investigate several of the approaches discussed above to the
maintainance of the $\divB=0$ constraint which are
applicable in an SPH context, namely the source term approach discussed in
the previous chapter (\S\ref{sec:monopoles}), projection methods
(\S\ref{sec:projection}) and the \citeauthor{dea02} approach (\S\ref{sec:hyperbolic}). The algorithm is then benchmarked, as
in the one dimensional case, against a wide range of multidimensional test
problems used to test recent grid-based MHD codes (\S\ref{sec:2Dtests}). The tests involve
the propagation of an initially non-zero magnetic divergence (\S\ref{sec:Bxpeaktest}), nonlinear Alfv\'en
waves (\S\ref{sec:alfven}), two dimensional shock tubes (\S\ref{sec:shock2D}), an MHD rotor problem (\S\ref{sec:mhdrotor}) and the Orszag-Tang vortex (\S\ref{sec:orstang}). The results are
summarised in \S\ref{sec:summary}.

%----------------------------------------------------------------------------------------------------------------
\section{Continuum equations}
\label{sec:continuum}
 Our SPMHD formalism is based on the equations of Magnetohydrodynamics in the form
\begin{eqnarray}
\frac{d\rho}{dt} & = & -\rho \pder{v^i}{x^i}, \label{eq:cty} \\
\frac{dv^i}{dt} & = & \frac{1}{\rho}\pder{S^{ij}}{x^j}, \label{eq:mom} \\
\frac{de}{dt} & = & -\frac{1}{\rho}\pder{(v_i S^{ij})}{x^j}, \label{eq:ener} \\
\frac{d}{dt}\left(\frac{B^i}{\rho} \right) & = & \frac{B^j}{\rho}\pder{v^i}{x^j}\label{eq:ind},
\end{eqnarray}
where 
\begin{eqnarray}
\frac{d}{dt} & = & \pder{}{t} + v^i \pder{}{x^i}, \nonumber \\
e & = & \frac12 v^2 + u + \frac12 \frac{B^2}{\mu_0\rho}, \\
S^{ij} & = & -P \delta^{ij}  + \frac{1}{\mu_0} \left( B^i B^j- \frac12 \delta^{ij} 
B^2 \right). \label{eq:stress}
\end{eqnarray}
 The formulation of these equations with respect to terms proportional to the
divergence of the magnetic field was discussed in paper~I and derived
self-consistently from a variational principle in paper~II. The implications of this
particular formulation of the continuum equations in the propagation of divergence errors is
discussed in \S\ref{sec:monopoles} and confirmed in the numerical tests presented
in \S\ref{sec:2Dtests}.

Note that in place of the
specific total energy $e$, the thermal energy can be evolved according to
\begin{equation}
\frac{du}{dt} = -\frac{P}{\rho}\pder{v^i}{x^i}.
\label{eq:utherm}
\end{equation}
 Similarly, in place of (\ref{eq:ind}) we could equivalently evolve the magnetic
flux density $B^i$ according to
\begin{equation}
\frac{dB^i}{dt} = -B^i \pder{v^j}{x^j} + B^j \pder{v^i}{x^j},
\label{eq:Bsrc}
\end{equation}

 The difference between evolving the total energy $e$ in place of the
thermal energy $u$ is found to be very minor. One
disadvantage of using the total energy is that it does not guarantee a positive thermal energy (although this
can be a useful diagnostic of when a simulation is going wrong). We choose
to evolve the magnetic flux per unit mass $B^i/\rho$ since it is the natural
variable to be carried by particles of fixed mass. Again, however, the difference
between using (\ref{eq:ind}) and (\ref{eq:Bsrc}) is found to be minor (although
some difference might be expected for simulations involving large changes in the
density). In general the differences between evolving different variables in SPH (and SPMHD) is dependent purely on the timestepping algorithm used and can be shown to decrease as smaller timesteps are used. It should be noted that these differences are much smaller than those found in grid-based codes due to the exact treatment of the advection terms in Lagrangian formulations.

The equation set is closed by an appropriate
equation of state, which for an ideal gas is given by
\begin{equation}
P = (\gamma -1) \rho u,
\end{equation}
where $P$ is the pressure, $u$ represents the internal energy per unit mass and
$\gamma$ is the ratio of specific heats.

%----------------------------------------------------------------------------------------------------------------
\section{SPMHD Equations}
\label{sec:spmhd}
 The discrete formulation of the SPMHD equations was discussed in paper~I and derived
self-consistently from a variational principle in paper~II. A self-consistent formulation of the
SPMHD equations in the case of a variable smoothing length was also derived in paper
II. We summarise the equations describing the method below.
 
 The continuity equation is expressed by the density summation
\begin{equation}
\rho_a = \sum_b m_b W(\vert\br_a-\br_b \vert, h_a),
\label{eq:rhosum}
\end{equation}
 where $W(\vert\br_a-\br_b \vert, h)$ is the interpolation kernel with
smoothing length $h$, for which we use the usual cubic spline \citep[paper
I,][]{monaghan92}. The time
derivative of (\ref{eq:rhosum}) gives the SPH expression for the continuity equation (\ref{eq:cty}), in the form
\begin{equation}
\frac{d\rho_a}{dt} = \frac{1}{\Omega_a} \smb \bv_{ab}\cdot\nabla_a W_{ab}(h_a),
\label{eq:sphcty}
\end{equation}
where $\bv_{ab} = \bv_{a} - \bv_{b}$ and $\Omega$ is a normalisation term which takes account of the variation of the smoothing
length with density (paper~II), given by\footnote{Beware that the expression given for $\Omega$
in paper~II contains some typographical errors. The correct expression is given by (\ref{eq:omegaa}).}
\begin{equation}
\Omega_a = \left[1 - \pder{h_a}{\rho_a}\smc
\pder{W_{ab}(h_a)}{h_a}\right].
\label{eq:omegaa}
\end{equation}
The smoothing length is assumed to depend on the density via the relation
\begin{equation}
h_a = \eta \left(\frac{m_{a}}{\rho_a} \right)^{1/\nu},
\label{eq:hrho}
\end{equation}
with derivative
\begin{equation}
\pder{h_a}{\rho_{a}} = -\frac{h_a}{\nu\rho_a},
\label{eq:dhdrho}
\end{equation}
where $\nu$ is the number of spatial dimensions. We enforce this relation by
calculating both $h$ and $\rho$ self-consistently by iteration of the density
summation (\ref{eq:rhosum}). The manner by which this is done is described in more
detail in paper~II and also in \citet{price04}. In brief, since
the density of each particle is independent of neighbouring particle smoothing
lengths we are able to iterate only those particles which have not converged. Furthermore
we predict the new value of the smoothing length using the time derivative
\begin{equation}
\frac{dh_a}{dt} = -\frac{h_a}{\nu\rho_a}\frac{d\rho}{dt}.
\label{eq:hevol}
\end{equation}
As a result the additional cost involved is minimal.

 The momentum equation (\ref{eq:mom}) in SPH form is given by
\begin{eqnarray}
\frac{dv^i_a}{dt} & = & \smb \left[ \left(\frac{S^{ij}}{\Omega\rho^2}\right)_a\pder{W_{ab}(h_a)}{x^j_a} \right.\nonumber \\
& & \left.+ \left(\frac{S^{ij}}{\Omega\rho^2}\right)_b
\pder{W_{ab}(h_b)}{x^j_a} \right], \label{eq:tensor}
\end{eqnarray}
 whilst the energy equation (\ref{eq:ener}) in discrete form is given by
\begin{eqnarray}
\frac{de_a}{dt} & = & \smb \left[ \left(\frac{S^{ij}}{\Omega\rho^2}\right)_a v^i_b
\nabla^j_a W_{ab}(h_a) \right. \nonumber \\
& & \left.+ \left(\frac{S^{ij}}{\Omega\rho^2}\right)_b v^i_a
\nabla^j_a W_{ab}(h_b) \right] \label{eq:spmhdener}.
\end{eqnarray}
The internal energy equation (\ref{eq:utherm}) in SPH form is given by
\begin{equation}
\frac{du_a}{dt} = \frac{P_a}{\Omega_a \rho_a^2} \smb \bv_{ab}\cdot\nabla_a
W_{ab}(h_a).
\label{eq:sphutherm}
\end{equation}

Finally, the induction equation is given by
\begin{equation}
\frac{d}{dt}\left(\frac{B^i}{\rho}\right)_a =
-\frac{B^j}{\Omega_a\rho_a^2}\smb v^i_{ab} \pder{W_{ab}(h_a)}{x^j_a}, \\
\label{eq:indsph}
\end{equation}
or alternatively
\begin{equation}
\frac{dB^i_a}{dt} = -\frac{1}{\Omega_a\rho_a}\smb
\left[v^i_{ab} B^j_a - B^i_a v^j_{ab}\right]\pder{W_{ab}(h_a)}{x^j_a}.
\label{eq:Bgradh}
\end{equation}

\section{Instability correction}
\label{sec:instability}
 In paper~I an artificial stress or `anticlumping' term described by \citet{monaghan97} was used to eliminate the
tensile instability associated with a conservative formulation of the momentum
equation in SPMHD. The basis of this approach is that the instability manifests as
particles clumping together in the presence of a negative stress and at short
wavelengths (see below). The solution proposed by \citet{monaghan97} and described in paper~I was to introduce a repulsive term proportional to the anisotropic magnetic force which acts to remove the instability at short wavelengths
by preventing particles from clumping together. This term has been used very
effectively in elastic dynamics simulations \citep{gms01} and was found to remove the
instability very effectively in the one dimensional simulations considered in papers I
and II. A more detailed investigation of the anticlumping term has been given recently
in \citet{price04}, interpreting the anticlumping term as a modification of the kernel
gradient used in the anisotropic force term. In this investigation several
disadvantages to the anticlumping approach were highlighted. The first is that at
large negative stresses (e.g. at low magnetic $\beta$) the anticlumping term can cause
the numerical estimate of the sound speed to be significantly in error. The second,
somewhat fatal disadvantage is that the anticlumping term does not appear to guarantee
stability in the case of a variable smoothing length. For more details we refer the
reader to \citet{price04}, however it suffices to say that
the anticlumping approach does not appear to be uniformly satisfactory for dynamical
MHD problems, particularly in more than one dimension. We therefore consider two
alternative approaches in this paper, which are outlined below (\S\ref{sec:subtractBconst},\ref{sec:otherposs}).

 It is worth recalling that the physical source of the instability is the additional
small but non-zero force directed parallel to the magnetic field in the conservative
formulation of the MHD equations (see introduction).
For this reason it might be expected that enforcing the $\divB = 0$ condition might
also eliminate the tensile instability. In fact this is not the case, since the
instability manifests even in one dimension (where the divergence is zero exactly).
The reason for this is that although the divergence is zero by virtue of $B_x =
\mathrm{const}$, the gradient of this constant (as evaluated in the force term) is not
necessarily zero numerically. In particular this is the case in the conservative
formulation of the SPMHD equations, since the symmetric SPH gradient evaluation in the form
\begin{equation}
\nabla A_a = \sum_b m_b \left( \frac{A_a}{\rho_a^2} + \frac{A_b}{\rho_b^2} \right)
\nabla W_{ab},
\end{equation}
is non-zero in the case of $A = \mathrm{const}$. To counter this problem two
approaches may be taken. The first is to add or subtract an arbitrary constant
in order to keep the total stress positive and thus preventing negative stresses (and
instability) from occurring. The second approach is to use an SPH gradient operator
which vanishes for constant functions. These approaches are described below.

\subsection{Removing the constant component of magnetic field}
\label{sec:subtractBconst}
  A simple method for removing the
tensile instability is to subtract an arbitrary constant from the stress in order to make the total stress positive. For simulations where the magnetic field is strong due to an initial net flux through the system, 
a natural choice for this constant is to subtract the external (ie. produced by
currents outside the simulation domain) component of the magnetic field. In this case the stress tensor (\ref{eq:stress}) for particle $a$ is modified according to
\begin{equation}
S^{ij}_a = -\left(P_a + \frac{1}{2\mu_0}B_a^2\right)\delta^{ij} +
\frac{1}{\mu_0}\left(B_a^i B_a^j - B^i_0 B^j_0\right),
\label{eq:stresssub}
\end{equation}
where $\bB_0$ is the magnetic field component which does not change throughout
the simulation (for example in one dimensional simulations we would use $\bB_0 = [B_x,0,0]$).
In general the constant field could also have a spatial profile (for example
in a fixed dipole field from the central star in an accretion disc) in which case the analytic gradient could be used. In all of the cases we consider the
external magnetic field is always the same independent of the particle
position, such that calculating (\ref{eq:stresssub}) involves storing only a single vector. It is worth noting that the formalism given above (where the constant field
is subtracted from the total field) is more efficient than explicitly adding
the contributions from separate constant and variable field components.

 This simple solution
completely cures the one dimensional instability because the $B_x$ component
of the field is explicitly removed from the anisotropic gradient term. Negative stresses can only arise in this formulation when the
anisotropic terms in the fluctuating component dominate the isotropic pressure
term (from which the constant field has \emph{not} been subtracted). 

A more general formulation which guarantees stability at all times must ensure that the total stress is positive. This can be achieved by using
\begin{equation}
S^{ij}_{a} = S^{ij}_{a} - S_{const},
\end{equation}
where
\begin{equation}
S_{const} = {\rm max} \left[ \left( \frac{1}{2} \frac{B^{2}}{\mu_{0}} - P \right), 0\right].
\end{equation}
where the maximum is taken over all the particles. In many ways this is similar to the original proposal of \citet{pm85} in which the maximum value of the stress tensor over all the particles was determined and
then subtracted from the stress for each particle. One disadvantage to this approach is that
total energy is not conserved exactly since the contribution to the total
energy evolution from the induction equation (which uses the total magnetic
field) does not exactly balance the contribution from the momentum equation.
An alternative is the approach of Morris (described below) in which the anisotropic term is modified slightly. In this paper we subtract the external field in simulations where a dominant external field is present, as in many of the two dimensional problems considered in this paper,
reverting to the Morris approach otherwise. In practice we find little to differentiate the two approaches. The results in all cases are much better than those obtained using the anticlumping term.

\subsection{Morris approach}
\label{sec:otherposs}
 An approach suggested by \citet{morrisphd} is to retain the
conservation of momentum on the isotropic terms in (\ref{eq:tensor}) but to treat the
anisotropic terms using a differencing formalism which is exact in the case of a
constant function (see above). The force term is then given by
\begin{eqnarray}
& - & \smb\left(\frac{P_a + \frac12 B^2_a/\mu_0}{\rho_a^2} + \frac{P_b + \frac12
B^2_b/\mu_0}{\rho_b^2}\right)\pder{W_{ab}}{x^i} \\
& + & \frac{1}{\mu_0}\smb
\frac{(B_iB_j)_b - (B_iB_j)_a}{\rho_a\rho_b}\pder{W_{ab}}{x_j}.
\label{eq:morrisforce}
\end{eqnarray}
This formalism does not therefore guarantee exact momentum conservation (since
the anisotropic term does not give equal and opposite forces between particle pairs) but can
be expected to give good results on shocks for which the anisotropic term is
less important. It is also a better approach than using formalisms
based on a pure ${\bf J}\times \bB$ force since (\ref{eq:morrisforce}) is still a discretisation of a tensor force and
therefore conserves momentum in the continuum limit for non-zero $\divB$. This
also means that (\ref{eq:morrisforce}) retains the consistent formulation of the MHD equations in the presence of monopoles,
although the discrete equations are no longer self-consistent with each other (where self-consistent means that the equations can be derived from a variational principle and will thus conserve momentum, energy and entropy). Note that when using the variable smoothing length terms, we use the average of
the normalised kernel gradient in (\ref{eq:morrisforce}), as in the dissipative
terms. The small amount of non-conservation introduced by the Morris formulation is not found to significantly affect the shock capturing ability of the scheme.

\section{Dissipative terms}
\label{sec:mhdav}
 Artificial dissipation terms which are required in order to simulate shocks were
formulated in paper~I. These terms are given by
\begin{eqnarray}
\left(\frac{d{\bf v}}{dt}\right)_{diss} & = & - \smb \frac { \alpha v_{sig} ({\bf v}_a -
{\bf v}_b ) \cdot \runit}{\bar{\rho}_{ab} } \nabla_a W_{ab},
\label{eq:vdiss} \\
\left(\frac{d\bB_a}{dt}\right)_{diss}  & = & \rho_a \sum_b m_b \frac{\alpha_{B} v_{sig}}{\bar{\rho}_{ab}^2}(\runit \times ( {\bf B}_{ab} \times \runit)r_{ab} F_{ab},
\label{eq:Bdiss} \\
\left(\frac{de_a}{dt}\right)_{diss} & = & - \smb \frac{v_{sig} (
e^*_a - e^*_b)}{\bar{\rho}_{ab} } \runit \cdot \nabla_a W_{ab},
\end{eqnarray}
where
\begin{equation}
\runit = \frac{(\br_a - \br_b)}{\vert \br_a - \br_b \vert}
\end{equation}
and
\begin{equation}
e^*_a = \left \lbrace \begin{array}{ll} \frac12 \alpha (\bv_a\cdot\runit)^2 + \alpha_u u_a \nonumber \\
\nonumber \\ + \frac12 \alpha_B [B_a^2 -
(\bB_a\cdot\runit)^2]/\mu_0\bar{\rho}_{ab}, & \bv_{ab}\cdot\br_{ab} < 0; \nonumber \\ \nonumber \\
 \alpha_u u_a + \frac12 \alpha_B [B_a^2 -
(\bB_a\cdot\runit)^2]/\mu_0\bar{\rho}_{ab}, & \bv_{ab}\cdot\br_{ab} \ge 0; \end{array} \right.
\nonumber
\end{equation}
with a similar expression for $e^*_b$.
 Note that the notation used in this paper differs slightly from that used in paper~I.
In particular we use separate parameters $\alpha$, $\alpha_u$ and $\alpha_B$ to
control the artificial viscosity, thermal conductivity and resistivity respectively
rather than a single parameter $K$. Note also that these parameters are expected to be
of order unity rather than $K\sim0.5$ (such that $\alpha$ corresponds to the $\alpha$ used in the
\citet{monaghan92} artificial viscosity formulation used widely in SPH).

 The signal velocity $v_{sig}$ represents the fastest speed of signal propagation between the two
particles. In MHD we use
\begin{equation}
v_{sig} = \frac12 \left[ v_a + v_b - \beta  {\bf v}_{ab} \cdot \runit \right],
\label{eq:vsigmag}
\end{equation}
where
\begin{eqnarray}
v_a & = & \frac12 \left (    \sqrt{ c_a^2 + \frac{B_a^2}{\rho_a \mu_0 } + \frac{2
{\bf B}_a \cdot \runit c_a}{ \sqrt{\rho_a \mu_0} } } \right. \nonumber \\ 
& & \left.+ \sqrt{ c_a^2 +
\frac{B_a^2}{\rho_a \mu_0 } -\frac{2 {\bf B}_a \cdot \runit  c_a}{ \sqrt{\rho_a
\mu_0} } }   \right ),
\end{eqnarray}
with a similar equation for $v_b$, where $c$ is the sound speed. Again our notation differs slightly from that used in paper~I as we use a dissipation parameter $\alpha$ of order unity and $v_{sig}\sim c_{s}$ (as opposed to $K\sim0.5$ and $v_{sig}\sim 2c_{s}$ in paper~I). The $\beta$ term in the signal velocity in this formalism naturally provides the non-linear (Von Neumann-Richtmyer) component of the artificial viscosity.

The dissipative terms (\ref{eq:vdiss}) and (\ref{eq:Bdiss}) provide an artificial
viscosity and resistivity. The term involving $(u_a - u_b)$ in the energy equation provides an
artificial thermal conductivity. These terms are derived so as to
guarantee a positive definite contribution to the thermal energy and entropy
(paper~I).
For reference the dissipative term added to the internal energy equation is given by
\begin{eqnarray}
\left(\frac{du}{dt}\right)_{diss} & = & -\smb \frac{v_{sig}}{\bar{\rho}_{ab}}\left\{ \frac{1}{2}\alpha \left[
(\bv_a\cdot\runit)
- (\bv_b\cdot\runit)\right]^2 \right. \nonumber \\
& & +  \alpha_u (u_a - u_b) \nonumber \\
& & + \left. \frac{\alpha_B}{2\mu_0\bar{\rho}_{ab}}\left[B_{ab}^2 - (\bB_{ab}\cdot\runit)^2\right]
\right\} \runit\cdot \gwab
\label{eq:udissmhd}
\end{eqnarray}

 In the one-dimensional tests described in paper~I, it was found that the
dissipative terms as described above were insufficient in shock problems
involving jumps in the transverse velocity component, resulting in a
modification of the viscosity term in order to apply the viscosity to all
velocity components. This was attributed to the fact that
the problems considered involved two and three dimensional velocity components 
whilst restricting the particles to move in only one spatial dimension. In the course of the multidimensional tests, it became apparent that this conclusion was incorrect. In paper~I, following the
usual procedure for hydrodynamical SPH, the dissipative terms were applied only
for approaching particles ($\bv_{ab}\cdot\runit < 0$). Whilst this is appropriate for the artificial
viscosity term, discontinuities in the magnetic field (requiring artificial
resistivity) can occur for particles in both compression and rarefaction. Applying the
artificial resistivity (\ref{eq:Bdiss}) uniformly across the simulation was found to correct the oscillations observed
in the magnetic field, and hence also in the transverse velocity components, removing
the need for any modification of the viscosity term. In fact the results with the
artificial resistivity term applied separately to the viscosity are
an improvement on those given in paper~I and are presented in \citet{price04}.

\subsection{Dissipation terms using total energy}
\label{sec:mhdavtoten}
 A further issue in a multidimensional context is that in the derivation of the above dissipative terms (paper~I) it was assumed that only components of the magnetic field
perpendicular to the line joining the particles would change at a shock front.
However, in a multidimensional simulation
the assumption of non-zero magnetic divergence may not hold exactly, as has
already been discussed. In particular divergence errors are often created at
flow discontinuities where fluid quantities are changing rapidly. It therefore
makes good sense to drop the assumption of non-zero magnetic divergence in the
derivation of the dissipative terms. The assumption that only the velocity
components parallel to the line joining the particles will change is also not strictly true in MHD since velocity components transverse to this line
will change with a jump in the transverse magnetic field. For this reason we
re-derive the dissipative terms with an energy term of the form
\begin{equation}
e^*_a = \frac{1}{2} \alpha \bv_a^2 + \alpha_u u_a + \alpha_B \frac{B_a^2}{2\mu_0\bar{\rho}_{ab}}
\end{equation}
which involves both the total kinetic and magnetic energies. The implication is that smoothing is then also applied to jumps in ${\bf B}$ parallel to the shock (ie. $\divB$ jumps) and via the kinetic term to transverse velocity jumps (ie. shear discontinuities). For the
contribution to the entropy to be positive definite, the terms in the thermal energy equation
must take the form
\begin{eqnarray}
\left(\frac{du}{dt}\right)_{diss} & = & -\smb
\frac{v_{sig}}{\bar{\rho}_{ab}}\left\{ \frac{1}{2}\alpha (\bv_a - \bv_b)^2 \right.
\nonumber \\
& & + \frac{\alpha_B}{2\mu_0\bar{\rho}_{ab}} (\bB_a - \bB_b)^2 \nonumber \\
& & \left. + \alpha_u (u_a -
u_b)\right\}\runit\cdot \gwab,
\end{eqnarray}
which correspondingly requires dissipation terms in the momentum and induction equations of the
form
\begin{eqnarray}
\left(\frac{d\bv_a}{dt} \right)_{diss} & = & \smb \frac{\alpha v_{sig} (\bv_a -
\bv_b)}{\bar{\rho}_{ab}} \runit\cdot\gawab,\label{eq:vdissfull} \\
\left(\frac{d\bB}{dt}\right)_{diss} & = & \rho_a \smb \frac{\alpha_B v_{sig}}{\bar{\rho}_{ab}^2}
\left(\bB_a -\bB_b \right)\runit\cdot\gwab \label{eq:Bdissfull}.
\end{eqnarray}
 In the multidimensional case we find that use of (\ref{eq:Bdissfull})
has distinct advantages over (\ref{eq:Bdiss}) since in more than one dimension
divergence errors can cause the extra component of the magnetic field to jump
slightly. Whether or not to use (\ref{eq:vdissfull}) in place of
(\ref{eq:vdiss}) is slightly less clear. The application of
dissipative terms to specific discontinuities in a hydrodynamic context is discussed in
\citet{price04} with regards to artificial thermal conductivity, where it
was found that smoothing of discontinuities in the thermal energy was necessary
only where the discontinuity is not already smoothed by the application of
artificial viscosity (which could occur, for example at a contact
discontinuity). In the present case, since a jump in transverse velocity can \emph{only} occur at a
corresponding jump in the transverse magnetic field, these discontinuities will
already be smoothed by the application of artificial resistivity there and so
the use of (\ref{eq:vdissfull}) may simply result in excessive dissipation (since
it must also be applied to particles in both compression and rarefaction,
whereas the usual viscosity term is applied only to particles in compression).
Furthermore (\ref{eq:vdissfull}) no longer conserves angular momentum (since the viscosity is not directed along the line joining the particles) and also no longer vanishes for rigid body rotation (since in effect rotational energy is converted into thermal energy). Thus
for simulations involving significant amounts of shear (for example in
accretion discs) the effects of using (\ref{eq:vdissfull}) would need to be
studied quite carefully. It is worth noting that a similar term was used by
\citet{morrisphd} for SPMHD shocks in place of an artificial resistivity.

\subsection{Dissipation switches}
\label{sec:avswitches}
 The artificial viscosity parameter $\alpha$ is controlled using the switch described in paper
I \citep{mm97}
\begin{equation}
\frac{d\alpha}{dt} = -\frac{\alpha-\alpha_{min}}{\tau} + \mathcal{S}
\end{equation}
 In paper~I we effectively used the source term
\begin{equation}
\mathcal{S} = \mathrm{max}(-2\nabla\cdot\bv,0.0)
\end{equation}
which (as noted in paper~I), is double the source term used by \citet{mm97} (note that in paper~I a dissipation parameter $K$ is used which has a value of $\alpha/2$). In
paper~I it was found that the stronger source term was necessary in order to
effectively damp post-shock oscillations. However for certain problems this source
term could cause the artificial viscosity to become overly strong, resulting in
excess smoothing of shock fronts. For this reason, in this paper we adopt a
modification of the switch proposed by \citet{rosswogetal00}, where the source term is
given by
\begin{equation}
\mathcal{S} = \mathrm{max}(-\nabla\cdot\bv,0.0)(2.0 - \alpha)
\end{equation}
resulting in an initially stronger source term which tails off as $\alpha$ reaches its
desired value of unity at the shock.

 Since artificial resistivity is required at discontinuities in the magnetic
field, which may occur where particles are not necessarily approaching each
other, artificial viscosity and resistivity should not be controlled using the
same switch, as was the case in paper~I, leading to unnecessary modifications of
the artificial viscosity term.
A similar switch appropriate to the artificial resistivity term can be devised
similar to that used in the viscosity. We evolve the resistive dissipation parameter
$\alpha_B$ according to
\begin{equation}
\frac{d\alpha_{B}}{dt} = -\frac{\alpha_B}{\tau} + \mathcal{S}
\label{eq:dalphaBdt}
\end{equation}
where in this case the source term is given by
\begin{equation}
\mathcal{S} = \mathrm{max}\left(\frac{\vert \nabla \times \bB
\vert}{\sqrt{\mu_0\rho}}, \frac{\vert \nabla \cdot \bB
\vert}{\sqrt{\mu_0\rho}} \right), 
\end{equation}
which has dimensions of inverse time, as required by (\ref{eq:dalphaBdt}).

 A similar switch may also be derived for the artificial thermal conductivity. A switch based on the first derivative of $u$ was used in \citet{price04}. In this paper we use a switch based on the second derivative of $u$, where the source term is given by
\begin{equation}
\mathcal{S} = 0.1 h\vert \nabla^2 u \vert, 
\end{equation}
where $h$ is the smoothing length and we multiply the source term by a small number in order to apply only the very minimum amount of dissipation needed to eliminate the wall heating effect. The second derivative term is computed according to \citep[e.g.][]{brookshaw85}
\begin{equation}
(\nabla^2 u)_{a} = 2  \smb \frac{(u_a - u_b)}{\rho_b } \frac{{\bf r}_{ab}\cdot
\gawab}{{\bf r}_{ab}^2},
\end{equation}
The second derivative switch is preferable since it responds only to sharp discontinuities in $u$, ensuring that a minimal amount of artificial thermal conductivity is applied. Additionally it requires storage of fewer quantities than the switch involving the first derivative. We have not investigated the use of switches for artificial viscosity or resistivity based on second (or higher) derivatives in this paper although it deserves further study. In particular a switch based on $\nabla (\nabla\cdot{\bf v})$ would be very useful in self-gravitating simulations where $-\nabla\cdot{\bf v}$ can have large constant values in the absence of shocks due to the gravitational collapse.

\section{Divergence correction techniques}
\label{sec:divb}

\subsection{Source term approach}
\label{sec:monopoles}
 The induction equation can be written in the `conservative' form
\begin{eqnarray}
\pder{\bB}{t} & = & -\nabla\times (\bv \times \bB), \label{eq:indcurl} \\
 & = & \nabla\cdot (\bv \bB - \bB \bv).
\end{eqnarray}
which explicitly conserves the volume integral of the flux
\begin{equation}
\int \bB \mathrm{dV}
\label{eq:fluxint}
\end{equation}
 In Lagrangian form (\ref{eq:indcurl}) can be written as
\begin{equation}
\frac{d\bB}{dt} = -\bB(\nabla\cdot\bv) + (\bB\cdot\nabla)\bv + \bv(\divB). 
\end{equation}
Taking the divergence of this equation, we have
\begin{equation}
\pder{}{t}(\divB) = 0,
\label{eq:divbevolcons}
\end{equation}
showing that the constraint $\divB = 0$ enters the MHD equations as an initial
condition. However allowing magnetic monopoles resulting from $\divB\neq 0$ to evolve
appropriately within the flow can prevent the build up of unphysical numerical effects
associated with their presence and can therefore reduce the need for computationally
expensive divergence cleaning procedures. Thus \citet{powell94} (see
\citealt{pea99}) suggested that the conservative forms of the MHD equations should
contain source terms to ensure that these errors are
propagated out by the flow. With this in mind, \citet{powell94} added
source terms to the momentum, energy and induction equations, which take the
(Lagrangian) form
\begin{eqnarray}
\frac{dv^i}{dt} & = & \frac{1}{\rho}\pder{S^{ij}}{x^j} -
\frac{B^i}{\rho}\pder{B^j}{x^j}, \label{eq:mompowell} \\
\frac{de}{dt} & = & -\frac{1}{\rho}\pder{(v_i S^{ij})}{x^j} -
\frac{v_i B^i}{\rho}\pder{B^j}{x^j},
\label{eq:enerpowell} \\
\frac{dB^i}{dt} & = & B^j\pder{v^i}{x^j} - B^i\pder{v^j}{x^j}.
\label{eq:indpowell}
\end{eqnarray}
 Taking the divergence of (\ref{eq:indpowell}) shows
that the divergence errors in this formalism evolve according to
\begin{equation}
\pder{}{t} (\divB) + \nabla\cdot (\bv \divB) = 0,
\label{eq:divBevol}
\end{equation}
which has the same form as the continuity equation for the density (where in
this case we have a density of magnetic monopoles, $\divB$). This therefore
implies that the total volume integral of $\divB$ across the simulation is
conserved and hence that the \emph{surface} integral of
the flux 
\begin{equation}
\int \bB\cdot \mathrm{dS} = \int (\nabla\cdot\bB) \mathrm{dV},
\label{eq:surfint}
\end{equation}
is conserved. The conservation of this quantity is far more important physically than the conservation of the volume integral
(\ref{eq:fluxint}).

 The disadvantage of using (\ref{eq:mompowell})-(\ref{eq:indpowell}) is that exact conservation of momentum and
energy is sacrificed, which proves to be important for shock-type problems.
Correspondingly it can lead to incorrect jump conditions at shock fronts
\citep{toth00}.  More recently it has been shown by \citet{janhunen00} and \citet{dellar01} that the
correct formulation of the MHD equations in the presence of monopoles should
\emph{not} violate the conservation of momentum and energy.

 The `monopole formulation' of \citet{janhunen00} and \citet{dellar01} is identical to the self-consistent formulation of the SPMHD equations derived in paper~II and given by (\ref{eq:cty})-(\ref{eq:ind}). Note that the induction equation (\ref{eq:Bsrc})
(equivalently using (\ref{eq:ind})) is the same as in Powell's 
method and therefore the same conclusions can be drawn regarding the manner in which the divergence
errors evolve (\ref{eq:divBevol}). We investigate the implications of the `source terms' in the induction equation in \S\ref{sec:Bxpeaktest}.

\subsection{Projection methods}
\label{sec:projection}
 A common approach to the divergence problem is to clean up the magnetic field
at regular intervals via the \emph{projection method} \citep[e.g.][]{bb80}. The basic
idea is to decompose the magnetic
field into a curl and a gradient (which can be done unambiguously for any vector field) according to
\begin{equation}
\bB^* = \curl \mathbf{A} + \nabla \phi.
\label{eq:projstep1}
\end{equation}
From this decomposition there are two ways of obtaining a divergence free field, both of which we discuss below.

\subsubsection{Scalar projection}
\label{sec:scalarproj}
Taking the divergence of this expression results in the Poisson equation
\begin{equation}
\nabla^2 \phi = \nabla \cdot \bB^*,
\label{eq:poisson}
\end{equation}
which can then be solved for the scalar quantity $\phi$. The magnetic field is
then corrected according to
\begin{equation}
\bB = \bB^* - \nabla \phi.
\label{eq:projstep3}
\end{equation}
 The major disadvantage with this approach is that the solution of the Poisson
equation (\ref{eq:poisson}) is computationally expensive, scaling as
$\mathcal{O}(N^2)$. In an astrophysical SPH context this may be offset somewhat by the fact that the
Poisson equation for the gravitational field is usually solved using a tree
code (e.g. \citealt{hk89,bea90}) which scales as $\mathcal{O}(N\mathrm{log}N)$. There
are, however, some subtleties to this approach, which we outline below.

 Projection schemes for incompressible flow in SPH have been implemented by
\citet{cr99}, the results of which are applicable to the present case. The important
point, also discussed by \citet{toth00} is that for the projection step to
reduce the divergence to zero (ie. to provide an \emph{exact} projection) requires that the discrete
version of (\ref{eq:poisson}) is satisfied exactly. This means that the operator
used to evaluate the divergence term on the right hand side of
(\ref{eq:poisson}) should be the same as the divergence operator used in the
evaluation of the $\nabla^2$ on the left hand side and that the gradient
operator used in the evaluation of $\nabla^2$ should be the same as that used in
(\ref{eq:projstep3}). \citet{cr99} approach this
problem by calculating the $\nabla^2$ using SPH operators, solving the Poisson
equation by matrix inversion. Good results were also obtained using an
approximate projection (ie. where the divergence operators on the left and
right hand side differ). In this scheme \citet{cr99} used the SPH evaluation of the Laplacian
similar to that which is commonly used for thermal conduction in SPH
\citep{brookshaw85,cm99} (and which is similar to the
artificial thermal conductivity term used in this paper). The Poisson equation is then solved by inverting the resulting matrix equation.

 The solution of (\ref{eq:poisson}) by direct summation (of which the tree code
is an approximation), uses the exact solution to the Poisson equation
(\ref{eq:poisson}) given by
\begin{equation}
\phi (\br) = \int G(\vert\br - \br' \vert) \divB (\br') \mathrm{dV} (\br'),
\end{equation}
where $G(\vert\br - \br' \vert)$ is the Green's function, given by
\begin{eqnarray}
G(r) & = & \frac{1}{2\pi} \ln {r} + \mathrm{const}, \nonumber \\
G(r) & = & -\frac{1}{4\pi r},
\end{eqnarray}
in two and three dimensions respectively. The gradient needed in the
correction step can be calculated directly, giving (in three dimensions)
\begin{equation}
\nabla\phi (\br) = -\frac{1}{4\pi}\int \frac{\divB (\br')}{\vert\br - \br' \vert^3} (\br - \br')
\mathrm{dV} (\br').
\end{equation}
 In SPH we replace the volume element $\rho \mathrm{dV}$ with the
mass per SPH particle and write the integral as a summation according to
\begin{equation}
\nabla\phi_a = -\smb \frac{(\divB)_b}{4\pi\rho_b} \frac{(\br_a - \br_b)}{\vert\br_a -
\br_b\vert^3}.
\label{eq:gradphi}
\end{equation}
 Since we still retain the freedom to choose the
discrete operator used to evaluate $\divB$ at each particle, it becomes clear
that the solution by direct summation will only provide an \emph{approximate}
projection, since (\ref{eq:poisson}) is not discretely satisfied. This
approximate solution will be degraded further when implemented using a tree
code. A further
disadvantage of the projection method for many of the problems considered in
this paper is that it is somewhat complicated to implement in the case of
periodic boundary conditions. The implication of these subtleties in the practical application of the projection method based on the Green's function solution (using a direct summation over the particles) are discussed in \S\ref{sec:Bxpeaktest}. Essentially we find that this projection method is reasonably effective at removing divergence errors at wavelengths larger than the smoothing length, but is less effective at removing short wavelength ($\sim h$) noise due to the smoothing of this noise inherent in the SPH operator used to calculate $\divB$. Preliminary calculations using this projection method in conjunction with a tree code in three dimensions indicate similar results. In this paper we compute the divergence of the magnetic field using the SPH operator
\begin{equation}
(\divB)_a = -\frac{1}{\Omega_{a}\rho_a}\smb (\bB_a - \bB_b)\cdot\gwab (h_{a}).
\label{eq:divBsph}
\end{equation}

\subsubsection{Vector projection}
\label{sec:vecproj}
 An alternative projection scheme can be implemented by solving for the vector
 potential $\bA$. That is,
we take the curl of (\ref{eq:projstep1}) to obtain
\begin{equation}
\nabla \times \mathbf{B^*} = \nabla(\nabla\cdot \bA) - \nabla^2 \bA.
\end{equation}
Choosing the Gauge condition $\nabla\cdot\bA = 0$, we obtain a Poisson equation for
the vector potential in terms of the current density $\mathbf{J} = \nabla\times{\bf
B^*}/\mu_0$
\begin{equation}
\nabla^2 \bA = -\mu_0 {\bf J}
\label{eq:poissonA}
\end{equation}
with solution
\begin{equation}
\bA (\br) = \int G(\vert\br - \br' \vert) {\bf J} (\br') \mathrm{dV} (\br').
\end{equation}
Taking the curl, we obtain an equation for the corrected magnetic field in terms of the
current density, which in three dimensions is given by
\begin{equation}
\bB = \nabla \times\bA = -\frac{\mu_0}{4\pi}\int \frac{{\bf J}(\br') \times (\br - \br')}{\vert\br - \br' \vert^3}
\mathrm{dV} (\br').
\label{eq:BfromJ}
\end{equation}
which is simply Biot-Savart's Law. In SPH form this is given by
\begin{equation}
\bB_a= - \smb \frac{(\nabla \times \bB^*)_b \times (\br_a -
\br_b)}{4\pi\rho_b\vert\br_a - \br_b \vert^3}.
\label{eq:BfromJsph}
\end{equation}
This method could also be useful in an SPH context in
situations where several disconnected
regions exist containing strong magnetic currents. By solving (\ref{eq:BfromJ}), the corrected magnetic field is determined from
the current density, resulting in a knowledge of the magnetic field at any point in
space. This approach was in fact used as the basis for the very first SPMHD algorithm
implemented by \citet{gm77}. In this paper we will only consider the use of \ref{eq:BfromJsph} as a divergence cleaning method. In this respect solving (\ref{eq:poissonA}) has a slightly higher computational expense than (\ref{eq:gradphi}) since the Poisson equation is
solved for a vector quantity rather than a scalar, giving (up to) three summations of $\mathcal{O}(N^{2})$ as opposed to just one. Nevertheless, there is a significant difference between the two methods. The difference is that whereas the approximate nature of the projection in (\ref{eq:gradphi}) means that the divergence is not guaranteed to be reduced to zero, in (\ref{eq:BfromJsph}) the divergence of the expression for $\bB$ \emph{is} zero exactly by virtue of the curl in the summation. 

The approximate nature of the projection in this case means that the current is not guaranteed to remain unchanged during the projection step. However \citep[as noted by][]{monaghan92} the current is usually well estimated by the SPH particles since the current is in general where the matter is. In this paper we compute the current density using the SPH operator
\begin{equation}
(\nabla\times\bB)_a = -\frac{1}{\Omega_{a}\rho_a}\smb (\bB_a - \bB_b)\times\gwab (h_{a}).
\label{eq:curlBsph}
\end{equation}

In practise we find that this projection method is far more effective than the scalar projection and this is demonstrated in \S\ref{sec:Bxpeaktest} (see Figure~\ref{fig:divbpeak_projection_r0}). Again preliminary three dimensional calculations indicate similar results, although an implementation using the tree code is more difficult in this case since all three components of the vector potential must be stored and summed over the tree as opposed to just one in the scalar projection (in which case the standard gravity tree can simply be called with the source term replacing the particle mass). However the degree to which the physical current is affected by this projection in three dimensions remains to be investigated.

\subsection{Hyperbolic divergence cleaning}
\label{sec:hyperbolic}
 \citet{dea02} examine alternative divergence cleaning procedures. In their
 paper \citep[see also][]{munzetal00}, they derive a general constrained formulation of the MHD
equations, from which formalisms can be derived to give divergence cleaning
which is elliptic (involving the
solution of a Poisson equation), parabolic (in which the divergence errors are
diffused away) and hyperbolic (where the divergence errors
are propagated away from their source at a characteristic speed). The projection method
described above is an elliptic approach, the main disadvantage to which is the
substantial computational cost involved in the solution of the Poisson equation. The parabolic approach was found to be
severely limited in scope due to the timestep restrictions imposed by the Courant
condition\footnote{an equivalent approach in SPMHD is to use an artificial
resistivity in order to diffuse away divergence errors. This has been used, for
example, by \citet{morrisphd} and \citet{hosking02}.}. The hyperbolic approach
was found to be particularly effective, especially when combined with a
parabolic term such that divergence errors are both transported and diffused.
It is this approach that we outline below in an SPH context.

 The basic idea is to introduce an additional scalar field $\psi$, which is
 coupled to the magnetic field by a gradient term in the induction equation,
\begin{equation}
\frac{d\bB}{dt} = -\bB(\nabla\cdot\bv) + (\bB\cdot\nabla)\bv -\nabla\psi.
\label{eq:indpsi}
\end{equation}
Note that our induction equation maintains the consistent treatment of divergence
terms discussed above. The variable $\psi$ is then calculated by adding an
additional constraint equation, which for the combined hyperbolic/parabolic
approach is given by
\begin{equation}
\frac{d\psi}{dt} = -c_h^2 (\divB) - \frac{\psi}{\tau}.
\label{eq:psievol}
\end{equation}
Neglecting the second term on the right hand side of (\ref{eq:psievol}) gives an
equation for $\psi$ which is purely hyperbolic. This implies that divergence
errors are propagated in a wave-like manner away from
their source with characteristic speed $c_h$ (for more details we refer the
reader to the \citeauthor{dea02} paper). The second term on the right hand
side is a parabolic term which causes $\psi$ to decay exponentially to zero with
e-folding time $\tau$ (this is easily seen by neglecting the hyperbolic term and
solving the resulting ordinary differential equation for $\psi(t)$). Since it is
desirable for the divergence errors to be propagated at the maximum possible
rate (within the timestep constraint imposed by the Courant condition), $c_h$
should be set equal to the maximum signal propagation speed. For simplicity we calculate
this as
\begin{equation}
c_h = \sqrt{\frac{\gamma P}{\rho} + \frac12 \frac{B^2}{\mu_0\rho}},
\end{equation}
where the maximum value over all of the particles is used.  The gradient term in the induction equation is calculated using a simple SPH estimate
\begin{equation}
\nabla \psi_a = \frac{1}{\Omega_{a}\rho_a} \smb (\psi_b - \psi_a) \gawab (h_{a}).
\end{equation}
Similarly the divergence of the magnetic field is calculated using (\ref{eq:divBsph}).

The choice of decay timescale $\tau$ is more complicated. In \citet{dea02} the decay timescale used is given by
\begin{equation}
\frac{1}{\tau} = \frac{c_{h}}{c_{r}} 
\end{equation}
 where they find that an optimal cleaning on their chosen test problem is given by $c_{r} = 0.1$. The problem with this is that $c_{r}$ is \emph{not} a dimensionless parameter, but rather has units of length. Thus the optimal choice for any given problem will depend on the length scales in that particular problem. We explicitly write the timescale as
 \begin{equation}
\frac{1}{\tau_a} = \frac{\sigma c_h}{\lambda_a},
\label{eq:decaytime}
\end{equation}
where $\lambda$ is a length scale and $\sigma$ is a dimensionless parameter which determines the decay timescale. Setting $\sigma = 0$ therefore gives a purely hyperbolic correction. The physical interpretation of the length scale in the problem can be determined by solving the following reduced system of equations (ie. neglecting the usual MHD evolution terms)
\begin{eqnarray}
\pder{\bB}{t} & = & -\nabla\psi \label{eq:divBclean1} \\
\pder{\psi}{t} & = & -c_h^2 (\divB) - \frac{\psi}{\tau}. \label{eq:divBclean2}
\end{eqnarray}
Taking the divergence of the (\ref{eq:divBclean1}) and substituting into (\ref{eq:divBclean2}) we obtain the following equation for $\psi$
\begin{equation}
\frac{1}{c_{h}^{2}}\pder{^{2} \psi}{t^{2}} - \nabla^{2} \psi + \frac{\sigma \lambda}{c_{h}} \pder{\psi}{t} = 0
\end{equation}
where an identical equation may be obtained for the evolution of $\divB$. This equation is simply the wave equation with a damping term, the solution to which is easily obtained by a separation of variables and is given in many standard textbooks. The length scale enters the solution as the wavelength for critical damping, that is the wavelength at which solutions change from being wave-like to being damped. 

 In practical calculations we expect divergence errors to be generated at wavelengths close to the smoothing length. We therefore set $\lambda = h$ and determine the value of the dimensionless parameter $\sigma$ by experiment. A value of
$\sigma = 0.2$ would imply that $\psi$ (and thus $\divB$) will have decayed significantly after the divergence errors have propagated approximately 5 smoothing lengths. In
\S\ref{sec:Bxpeaktest} we examine in detail the effects the hyperbolic cleaning on a problem involving a fixed wavelength of error (ie. independent of $h$) which graphically illustrates the divergence cleaning method (see Figure~\ref{fig:divbpeak1}). The effect of this type of cleaning on errors generated by the flow are examined in \S\ref{sec:orstang}. We find that values of $\sigma \sim 0.4-0.8$ generally give the best results, giving a good balance between the hyperbolic (fast but
non-diffusive) and parabolic (diffusive but slow-acting) effects. In practise
some diffusion is also added by the artificial resistivity terms (\S\ref{sec:mhdav}). In general, however, the divergence correction provided by the hyperbolic/parabolic scheme is found to be quite small (around a factor of $\sim 2$ reduction). Thus, whilst this type of divergence cleaning essentially comes free-of-charge computationally, under some circumstances it may be necessary to supplement it with a stronger form of cleaning, such as use of a projection method or some other kind of elliptic or parabolic cleaning which is not limited to the explicit time step condition. 

\section{Numerical tests}
\label{sec:2Dtests}
 The main issue to be addressed in 2D and 3D problems is the non-zero
divergence of the magnetic field. In the SPH context it also allows us to
estimate the extent to which the artificial dissipation spuriously affects the numerical
results. Again there is a substantial literature of multi-dimensional MHD problems
which have been used to test grid-based MHD codes (e.g. \citealt{dw94,rea95,balsara98,dw98,toth00}) and we consider several of these problems
here.

\subsection{Implementation}
 The implementation of the SPMHD equations used for the multidimensional tests
is almost identical to that used in the one dimensional case
(paper~I). The density is calculated by summation, the total energy
equation is used (although results are indistinguishable using the thermal energy
equation in nearly all cases) and the magnetic field is evolved using (\ref{eq:indsph}) (or
using (\ref{eq:indpsi}) when using the hyperbolic cleaning). In the shock tube tests we use unsmoothed initial
conditions. The artificial dissipative terms, except where otherwise indicated
are implemented using the jump in total magnetic energy (\S\ref{sec:mhdavtoten}) but the viscosity term uses only the velocity component along the line
joining the particles (\ref{eq:vdiss}). Unless otherwise indicated,  artificial viscosity and thermal
conductivity are applied using the switches discussed in \S\ref{sec:avswitches}
whilst the artificial resistivity term is applied uniformly using $\alpha_B =
1$. A major difference between the simulations presented here and those in the
paper~I is that the anticlumping approach was not found to be
uniformly successful in eliminating the tensile instability for all of the problems
considered (in particular for the Alfv\'en wave test only a narrow range of parameters
would produce stable results). Furthermore this term was found to result in spurious extra numerical noise, particularly in the shock tube tests. For this reason we have eliminated the tensile
instability by simply subtracting the constant component of the
magnetic field from the gradient term (\S\ref{sec:subtractBconst}) in the shock tube problems and using the stable Morris formulation of the magnetic force
(\S\ref{sec:otherposs}) elsewhere. Note that even on the shock problems the differences in results between these two methods is almost negligible.

\subsubsection{Error estimates}
 Various estimates can be made of the error produced in the simulation by any
non-zero magnetic divergence. Monitoring these quantities over the course of a
simulation thereby gives some measure of the magnitude of the error produced by
$\divB$. The most common approach in SPH implementations to date has been to
monitor the dimensionless quantity
\begin{equation}
\frac{h \divB}{\vert\bB\vert}
\end{equation}
and ensure that it remains small (typically $< 0.01$) over most of the simulation, where $h$ is the SPH smoothing
length and the divergence is calculated using (\ref{eq:divBsph}). This
provides some measure of the relative error in the magnetic field but no
indication of how much influence this error has in the dynamics. For this reason
it is also useful to measure the relative error in the total force caused by a non-zero divergence,
\begin{equation}
E_{force} = \frac{\mathbf{f}_{mag} \cdot \bB}{\vert\mathbf{f} \vert\vert \bB\vert}
\end{equation}
where $\mathbf{f}_{mag}$ is the magnetic component of the SPH force (\ref{eq:tensor}), whilst
${\bf f}$ is the total force on the particle. It is also useful to simply monitor the
evolution in the maximum, minimum and average of $\vert\divB\vert$ with time
as well as various conserved quantities.

\subsubsection{Conserved quantities}
\label{sec:conservedmhd}
 Aside from the usual conserved quantities of mass, momentum,
angular momentum, energy and centre of mass, several additional
quantities can be measured in MHD which can be useful diagnostics in a numerical
simulation. A list of such quantities can be derived
using Hamiltonian techniques and is given by (e.g.) \citet{mh84}. The helicity,
\begin{equation}
\int (\bA \cdot \bB) \mathrm{dV},
\end{equation}
where $\bB = \nabla\times\bA$, is a measure of the linkage of magnetic field lines (expressing the fact
that magnetic field lines which are initially linked cannot become unlinked in
the absence of dissipative terms). This quantity can only be usefully measured in
simulations which explicitly use the vector potential $\bA$. A similar invariant is the
cross helicity
\begin{equation}
\int (\bB \cdot \bv) \mathrm{dV} \approx \smb \frac{\bB_b}{\rho_b}\cdot \bv_b,
\label{eq:crosshel}
\end{equation}
 which measures the mutual linkage of magnetic field and vortex lines. The
conservation of the cross helicity is a result of the magnetic field lines being
frozen into the fluid. Measurement of the conservation of this quantity in a numerical
simulation therefore provides an estimate of the degree of slippage of the
magnetic field lines through the fluid. The volume integral of the magnetic flux
(\ref{eq:fluxint}) is also conserved across the simulation volume, provided that the flux is normal
to (or zero at) the boundary of the integration volume. However the conservation
of flux in a volume sense is not particularly important physically
\citep{janhunen00}. More important is that the surface integral of the flux
(\ref{eq:surfint}) should be conserved. The conservation of these quantities with
respect to formulations of the MHD equations in the presence of magnetic monopoles was
discussed in \S\ref{sec:monopoles}.

 There is also a conserved quantity which is the MHD analogue of the circulation
\citep{bo00,kr00}, although the physical interpretation is somewhat obscure. It has
been shown that SPH conserves an approximate version of the circulation in the
hydrodynamic case \citep{mp01}, related to the invariance of the equations to
the relabelling of particles around a closed loop due to the frozen-in vorticity field. A similar, though
more restricted relabelling symmetry holds in the MHD case (in that the particles around
the loop must also be on the same field line) and it may therefore be expected
that SPMHD also maintains this invariance. 

\subsubsection{Visualisation}
 In order to make a direct comparison of our results with those of grid-based MHD
codes, we interpolate the results from the particles to an array of pixels using
the SPH kernel. That is, for a contour or rendered plot of a scalar quantity $\phi$ we
interpolate to the pixels using
\begin{equation}
\phi(x,y) = \sum_b m_b \frac{\phi_b}{\rho_b} W(x - x_b, y-y_b, h_b)
\end{equation}
where $W$ is the cubic spline kernel used in the calculations \citep[paper~I;][]{monaghan92} and the summation is over
contributing particles. Note that in practise this is quite simple to implement, as it
involves only one loop over the
particles, during which the contributions from the current particle to all pixels within a smoothing
radius ($2h$) are calculated. For a vector quantity a similar interpolation can be
performed for each component. An interactive plotting program incorporating these interpolation schemes for visualisation of SPH (and SPMHD) data in 1, 2 and 3 dimensions has been written by the author and is available upon request.

\subsection{$\nabla\cdot \bB$ advection}
\label{sec:Bxpeaktest}
 The first problem we examine is a simple test similar to that used by \citet{dea02} in which a non-zero
magnetic divergence is introduced into the simulation as an initial
condition. This is a particularly good test for comparing various divergence
cleaning procedures. The initial conditions are a uniform density distribution
($\rho = 1$) in the domain $-0.5 < x < 1.5, -0.5 < y < 1.5$ with a constant
initial velocity field $\bv = [1,1]$. The initial gas pressure is $P=6$ with $\gamma=5/3$ and the magnetic field has
a constant component perpendicular to the plane $B_z = 1/\sqrt{4\pi}$. The
divergence is introduced as a peak in the $x-$component of the field in the form
\begin{equation}
B_x = (r/r_{0})^8 - 2(r/r_{0})^4 + 1 \hspace{1.5cm} r = \sqrt{x^2 + y^2}
\label{eq:Bxinit}
\end{equation}
where $r_{0}$ is the radius of the initial peak. The setup used here differs from that used by \citet{dea02} in that $r_{0}$ is a changeable parameter (using $r_{0}=1/\sqrt{8}$ gives their setup). The reason for this is that we find that the effectiveness of the divergence cleaning strongly depends on the wavelength of the divergence errors. Testing divergence cleaning procedures based on a single wavelength of error \citep[as in][]{dea02} can lead to misleading interpretations and an incorrect choice of parameters when applied to divergence errors which are generated in the course of real simulations.

\begin{figure}
\begin{center}
\begin{turn}{270}\epsfig{file=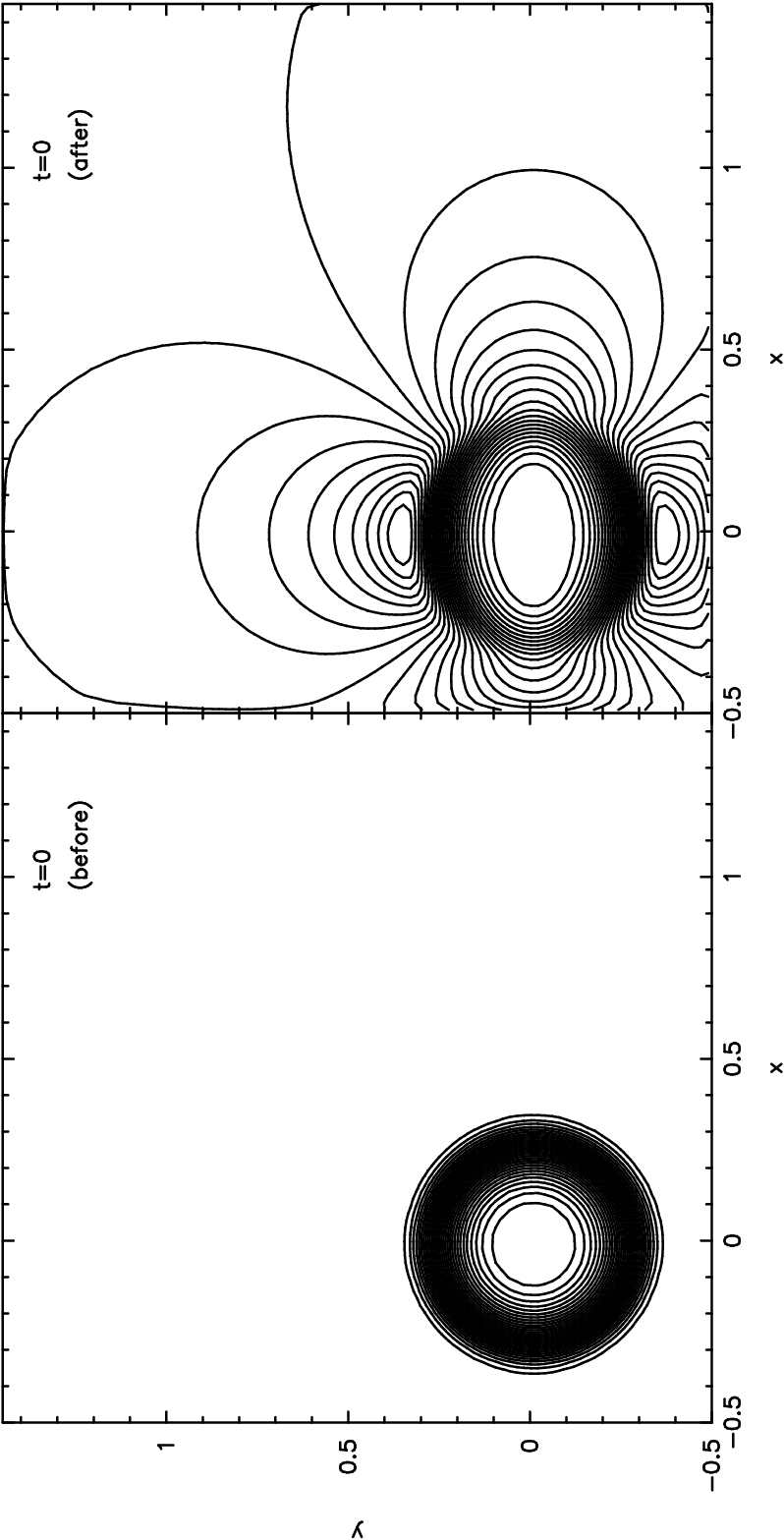,height=\columnwidth}\end{turn}
\caption{Divergence cleaning using the approximate (scalar) projection method described
in \S\ref{sec:projection}. The plot shows 30 contours of $B_x$ in the $\divB$ advection problem before (left) and after (right) a
single projection step at $t=0$.  The projected magnetic field adopts an
essentially divergence-free configuration in a single step. Note that the wavelength of the initial divergence error in this case is substantially larger than the smoothing length.}
\label{fig:divbpeak_projection}

\begin{turn}{270}\epsfig{file=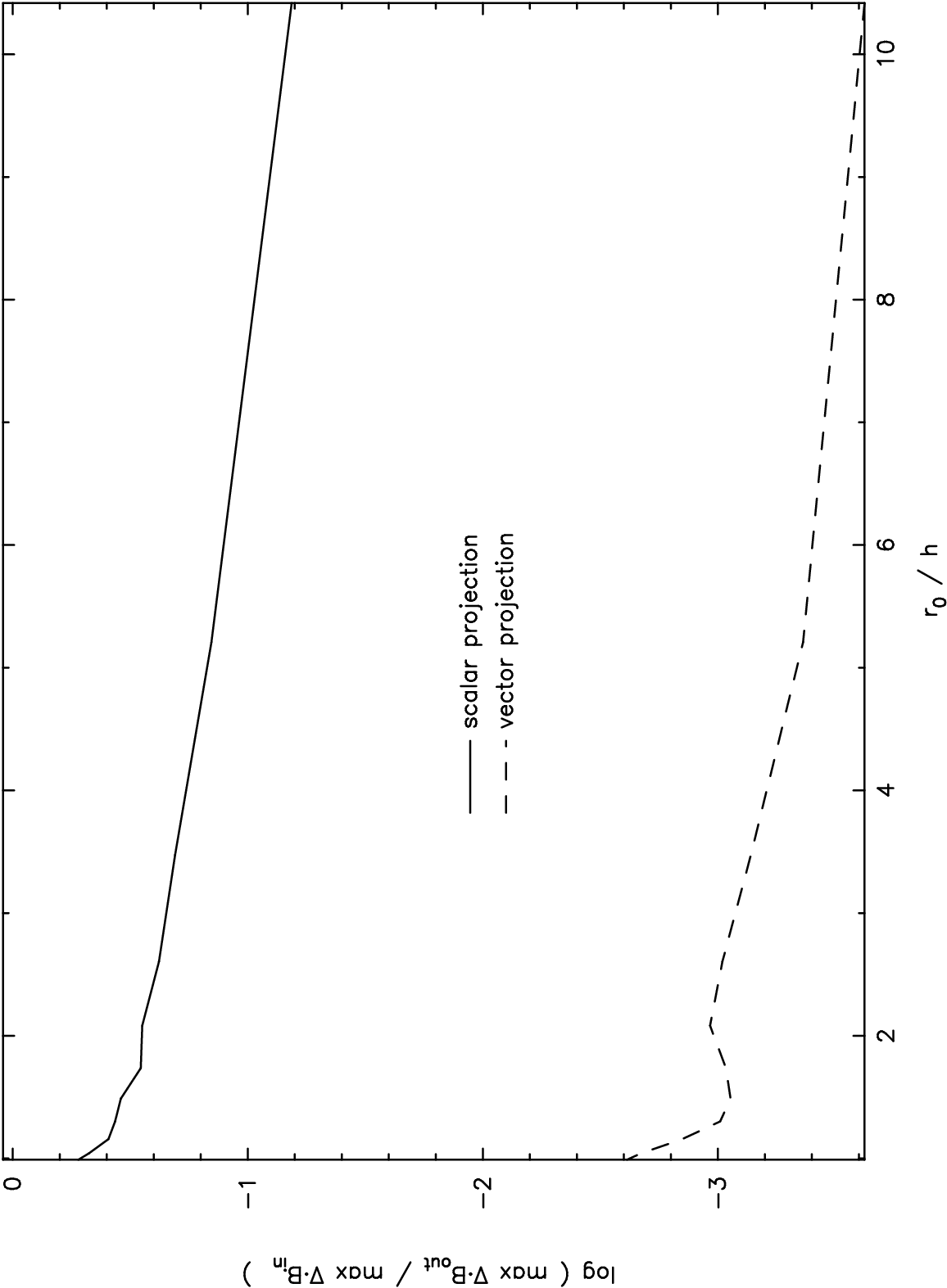,height=\columnwidth}\end{turn}
\caption{Wavelength dependence of the scalar and vector approximate projection methods described in \S\ref{sec:projection}. The $y-$axis shows the relative change in the maximum divergence error (ie. $\mathrm{max}(\divB_{new}/\divB_{0})$ over a single projection step taken at $t=0$. The $x-$axis gives the radius of the initial peak in the $x$ component of the field $r_{0}$ in units of the smoothing length $h$. In both cases the projection step becomes less effective as the wavelength of the divergence error approaches the smoothing length (ie. $r_{0}/h \to 1$), however the vector projection outperforms the scalar projection by a factor of $\sim 100$ at all wavelengths.}
\label{fig:divbpeak_projection_r0}
\end{center}
\end{figure}

 The contours of the initial $B_{x}$ field arrangement (\ref{eq:Bxinit}) in the \citet{dea02} case of $r_{0} = 1/\sqrt{8}$ are shown in the left hand side of Figure~\ref{fig:divbpeak_projection} (and similarly in Figure~\ref{fig:divbpeak1}). The particles are arranged on a cubic lattice for simplicity and in the evolution calculations the periodic boundary conditions are enforced using ghost particles.
Since the density is uniform throughout the simulation the results are
insensitive to whether ${\bf B}$ or ${\bf B}/\rho$ is evolved and
also to the instability correction method since the simulation is not unstable to negative stress. The artificial dissipation terms are turned off for this problem in order to isolate the effects of the divergence cleaning procedures. 

\subsubsection{Projection methods}
 We use this problem to test the projection methods by applying a single projection step to the divergence error introduced in the initial conditions. To illustrate the divergence-free configuration adopted by the field we plot in Figure~\ref{fig:divbpeak_projection} the contours of $B_{x}$ before and after the projection step. The initial configuration (left panel) is set according to (\ref{eq:Bxinit}) with $r_{0} = 1/\sqrt{8}$, corresponding to that used by \citet{dea02}. The right panel shows the resulting field configuration after the projection step. We have used 2,500 ($50\times50$) particles in this case, setup on a cubic lattice with a smoothing length set using $\eta = 1.2$ in (\ref{eq:hrho}), giving $h=0.048$. This means that the wavelength of the initial divergence error is substantially ($\sim 7\times$) larger than the smoothing length. However, we expect that divergence errors generated in the course of real simulations will tend to have wavelengths closer to $\sim h$. For this reason we have extended the test problem of \citet{dea02} to a variety of wavelengths by varying the parameter $r_{0}$. 
 
 The relative divergence cleaning given from a single projection step using both the scalar (\S\ref{sec:scalarproj}) and vector (\S\ref{sec:vecproj}) projection methods are plotted against $r_{0}/h$ in Figure~\ref{fig:divbpeak_projection_r0}. The particles have been setup as previously, in this case varying $r_{0}$ whilst keeping $h$ fixed. We have also performed simulations where $r_{0}$ is fixed and $h$ is varied by changing the number of particles. The results in both cases are virtually identical. The results are also insensitive to the absolute size of the divergence error, since we plot only the relative change in $\divB$. 
 
 For simplicity we have assumed that the boundaries are open when calculating the sum in the projection step, rather than explicitly treating the periodic boundaries. This is a reasonable assumption whilst the source terms for the Poisson equation (ie. $\divB$ or $\nabla\times\bB$) are non-zero in only a finite region of the simulation volume. However, to ensure that the maximum $\divB$ value is not an artefact of our non-treatment of the periodic boundary conditions, the divergence near the boundaries of the domain (on particles within $2h$ of the boundary) has been set to zero when calculating the maximum used in Figure~\ref{fig:divbpeak_projection_r0}. These calculations shown in Figure~\ref{fig:divbpeak_projection_r0} have also been performed with the $B_{x}$ peak placed in the centre of the domain rather than at the origin in order to move the source further away from the boundaries.
 
  It can be seen from Figure~\ref{fig:divbpeak_projection_r0} that the effectiveness of the divergence cleaning given by the scalar projection is reduced as the wavelength of the divergence error approaches the smoothing length (for the scalar projection the reduction in error approaches a mere factor of $\sim 2$ as $r_{0}/h \to 1$). Whilst the vector projection (dashed line) shows a similar trend with wavelength, it is clear that this method is a vast improvement over the scalar projection, reducing the maximum divergence error by a factor of $\sim 100$ compared to the scalar case. This is due to the fact that the vector projection gives an expression for $\bB$ which analytically has a zero divergence (refer to the discussion in \S\ref{sec:vecproj}). The only non-zero divergence resulting in this case is due to the SPH operator used to compute $\divB$ after the projection step. This is therefore a substantial improvement over the scalar projection, in which $\divB$ is not guaranteed to be exactly zero in \emph{any} approximation. We have also experimented with the scalar projection using SPH operators for $\divB$ other than (\ref{eq:divBsph}), all of which we find show similar results.
  
\subsubsection{Source terms and Hyperbolic/parabolic cleaning}

\begin{figure*}
\begin{center}
\epsfig{file=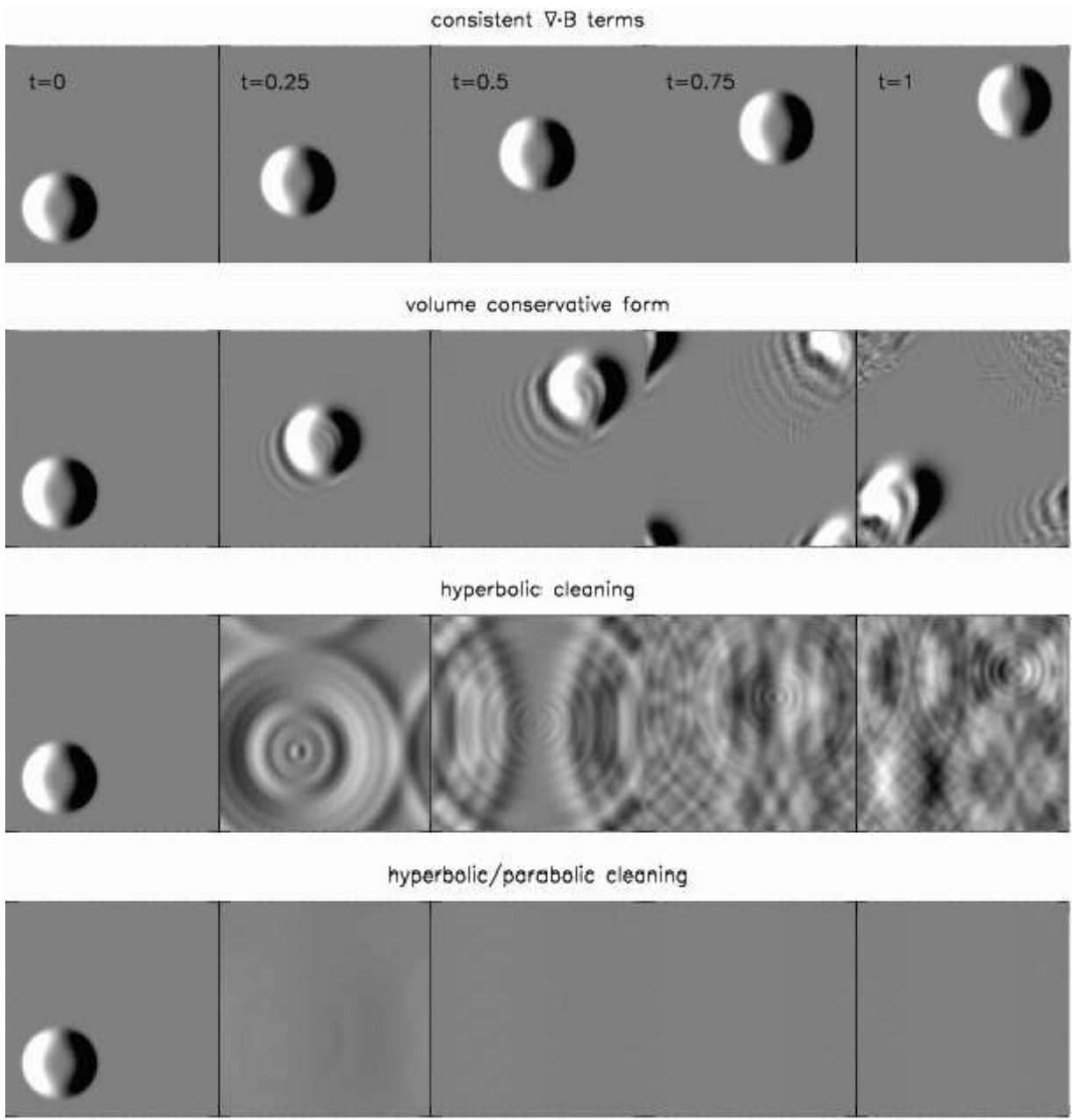,height=0.95\textwidth}
\caption{Results of the $\divB$ advection problem. An initially non-zero divergence is setup as a
peak in the $x-$component of the magnetic field (leftmost figures), with a
velocity field $\bv(x,y) = [1,1]$ and periodic boundaries. The plots show renderings of $\divB$ in the range $-1 < \divB < 1$ (from black to white) at various times throughout the simulation for various divergence cleaning
procedures. The consistent treatment of $\divB$ terms (top row) is clearly
seen to advect the divergence without change, which is an improvement over a ``conservative''
formulation of the MHD equations in which the divergence is smeared throughout the simulation
volume (second row). With the use of hyperbolic
cleaning in addition to the consistent $\divB$ terms, the divergence error is spread rapidly in a wavelike manner (third row), whilst with a mixed
hyperbolic/parabolic cleaning (fourth row) this error is also
quickly diffused away. }
\label{fig:divbpeak1}
\end{center}
\end{figure*}

 The source term approach (\S\ref{sec:monopoles}) and the hyperbolic/parabolic cleaning (\S\ref{sec:hyperbolic}) are tested by evolving the initial field configuration given by (\ref{eq:Bxinit}) forward in time. The results of this test 
are shown in Figure \ref{fig:divbpeak1}. The plots show the divergence of the magnetic field as it evolves in each case. The results using the consistent
formulation of $\divB$ terms discussed in paper~I and in
\S\ref{sec:monopoles} are shown in the top row. In this case the divergence
error is passively advected by the flow and both the field and the divergence error
remain unchanged (relative to the flow) at $t=1$, demonstrating that the formalism is indeed consistent
in the presence of magnetic monopoles and conserves the integral (\ref{eq:surfint}). 

In order to compare these results with a
conservative [in the sense of conserving (\ref{eq:fluxint})] formulation of the MHD equations, we have performed a simulation
using an SPH induction equation of the
form
\begin{equation}
\frac{d}{dt}\left(\frac{B^i_a}{\rho_a}\right) =
 \smb \left[ \frac{B_a^j}{\rho_a^2}  (v_b^i - v_a^i) 
 + \frac{v_a^i}{\rho_a^2} (B_b^j - B_a^j) \right] \pder{W_{ab}}{x^j_a}
\label{eq:indconssph}
\end{equation}
which is an SPH form of the conservative (in a volume sense) induction equation
\begin{equation}
\frac{d}{dt}\left(\frac{\bB}{\rho}\right) =
\left(\frac{\bB}{\rho}\cdot\nabla\right)\bv + \bv
\left(\frac{\divB}{\rho}\right).
\label{eq:indcons}
\end{equation}
The results using this formalism are shown in the second row of Figure
\ref{fig:divbpeak1}. The peak in $B_x$ is distorted by the flow and
the divergence error is smeared throughout the simulation. 

The third row in Figure
\ref{fig:divbpeak1} shows the results using the divergence correction discussed in
\S\ref{sec:hyperbolic} using only the hyperbolic term in (\ref{eq:psievol})(ie.
with $\sigma = 0$) in conjunction with the usual monopole formulation of the
induction equation (\ref{eq:indsph}). The divergence error is spread rapidly
in a wavelike manner by the constraint equation. However, the magnitude does not decrease substantially in this case.

 Using the mixed hyperbolic/parabolic cleaning with a small amount of diffusion (using the parabolic term in (\ref{eq:psievol}), in this case with $\sigma = 0.1$), this error is rapidly diffused away, resulting in a divergence-free field configuration (Figure \ref{fig:divbpeak1}, bottom row). For
comparison, the results of a single projection step at $t=0$ are shown in Figure
\ref{fig:divbpeak_projection}, showing the divergence-free configuration adopted
by the field.

\begin{figure}
\begin{center}
\begin{turn}{0}\epsfig{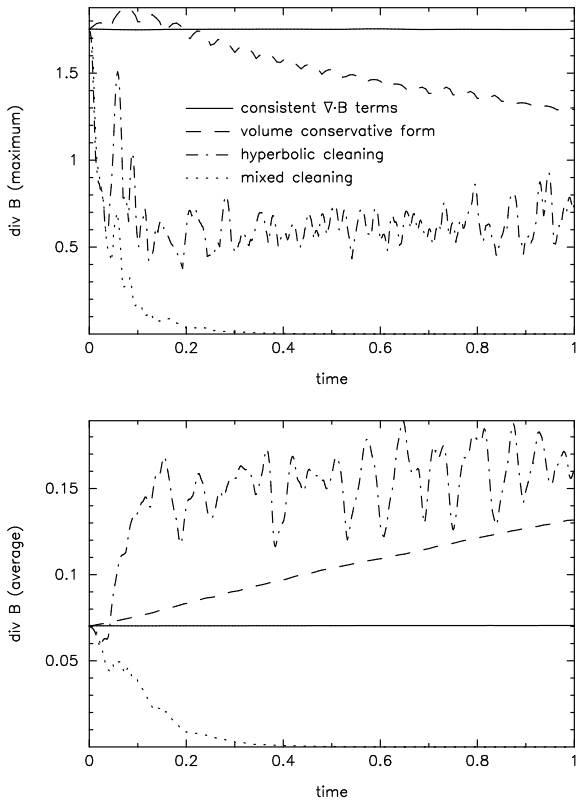}\end{turn}
\caption{Time evolution of the maximum (left) and average (right) values of $\vert \divB \vert$ over
the particles. With a conservative formulation of the induction equation the
divergence error increases with time (dashed line) whereas the errors are
conserved using a formulation which is consistent in the presence of magnetic
monopoles (solid line). With hyperbolic
cleaning (dot-dashed) the maximum is quickly reduced although the average
increases, however with the parabolic term included the
average error is also rapidly diffused away (dotted line).}
\label{fig:divbschemes_time}
\end{center}
\end{figure}

 The time evolution of various quantities throughout these simulations are shown
in Figure \ref{fig:divbschemes_time}. The left panels show the evolution of the maximum (top) and
average (bottom) of $\vert\divB\vert$. In conservative form (solid line) the maximum
divergence varies slightly and initially becomes larger than the initial value.
The bottom panel shows that the average value in this case steadily increases
over time, due to the smearing effect of the divergence propagation
(\ref{eq:divbevolcons}). The consistent formulation of $\divB$ terms (dashed
line) maintains a steady value of both the maximum and average, as observed in
Figure \ref{fig:divbpeak1}. With hyperbolic cleaning (dot-dashed) the maximum
divergence error is quickly reduced (although increases at later times as the
divergence waves cross the periodic domain and interact) whilst the average
climbs as the divergence error is spread throughout the domain. Using the mixed
hyperbolic/parabolic cleaning as described above (dotted line), both the maximum and average
divergence is swiftly reduced.

 Finally it is important to examine the effect of varying the strength of the parabolic (diffusion) term in
(\ref{eq:psievol}). The effects of varying the diffusion parameter $\sigma$ for this particular problem were explored in \citet{price04}. It was later realised however that these results depended strongly on the value of the smoothing length (ie. the resolution of the calculations). Despite this the optimal parameter in other simulations was found to be independent of $h$. The reason for this is quite simple in hindsight and is due to our use of $h$ as the length scale (wavelength of critical damping) in (\ref{eq:decaytime}). In this test problem the divergence error is setup in the initial conditions at a wavelength $r_{0}$ which is independent of the actual resolution used. Thus using a length scale $h$ in (\ref{eq:decaytime}), the optimal cleaning is strongly dependent on resolution, since the optimal wavelength in this case corresponds closely to the wavelength of the initial divergence error.  This also explains why \citet{dea02} found that their use of a fixed parameter $c_{r}$ which has dimensions of length was found to give cleaning which is independent of resolution (but only for this specific problem!). Since in realistic calculations divergence errors are produced at wavelengths $\sim h$, in general the length scale used in (\ref{eq:decaytime}) should reflect this. We therefore retain the length scale $h$ but defer examination of the effect of varying the parameter $\sigma$ to the Orszag-Tang vortex problem (\S\ref{sec:orstang}) where divergence errors are generated in the course of the evolution.

\subsection{Circularly polarized Alfv\'en wave}
\label{sec:alfven}

\begin{figure}
\begin{center}
\begin{minipage}{0.44\columnwidth}
\begin{turn}{270}\epsfig{file=malfven_setup.ps,height=\textwidth}\end{turn}
\end{minipage}
\hspace{0.02\columnwidth}
\begin{minipage}{0.5\columnwidth}
\epsfig{file=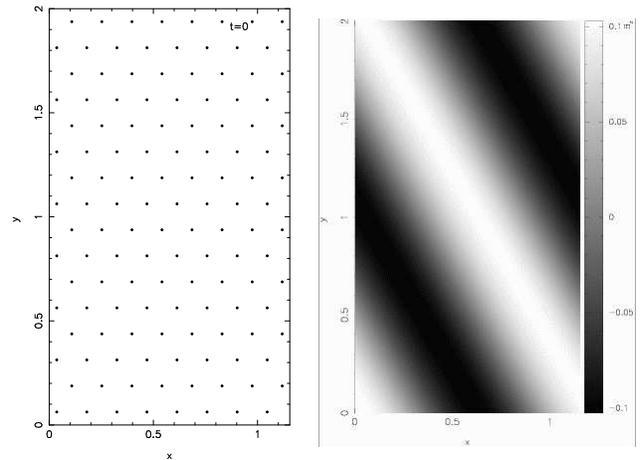,width=\textwidth}
\end{minipage}
\caption{Circularly polarized Alfv\'en wave test. The left figure shows the particle
setup in the lowest resolution simulation. On the right the vertical component of the magnetic
field is plotted as a rendered image from the $32\times 64$ particle run at $t=5$, showing the propagation of the wave with
respect to the domain and the particle setup.}
\label{fig:malfvensetup}
\end{center}
\end{figure}

\begin{figure*}
\begin{center}
\begin{minipage}{0.4\textwidth}
\begin{turn}{0}\epsfig{file=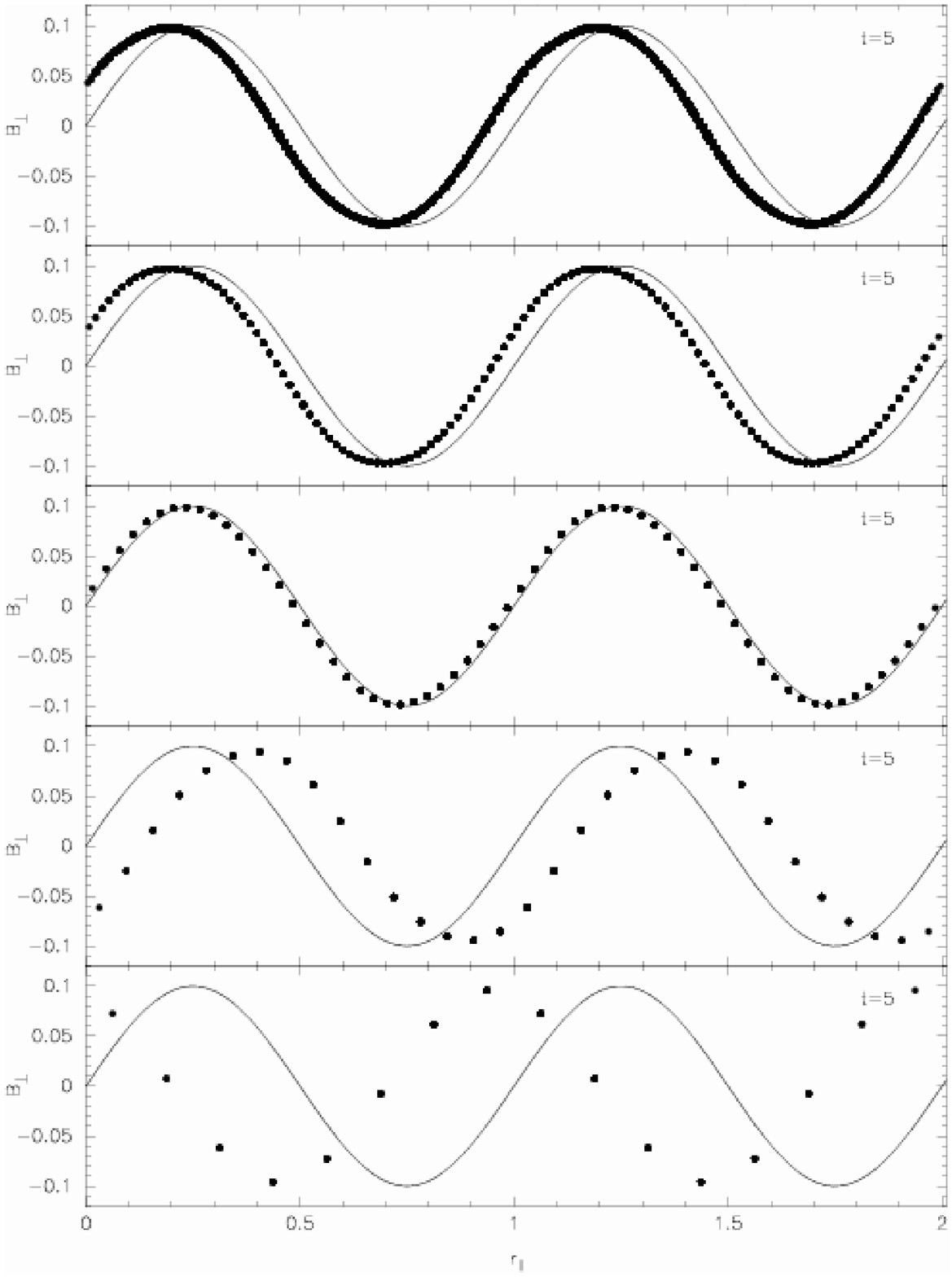,width=\textwidth}\end{turn}
\end{minipage}
\hspace{0.01\textwidth}
\begin{minipage}{0.4\textwidth}
\begin{turn}{0}\epsfig{file=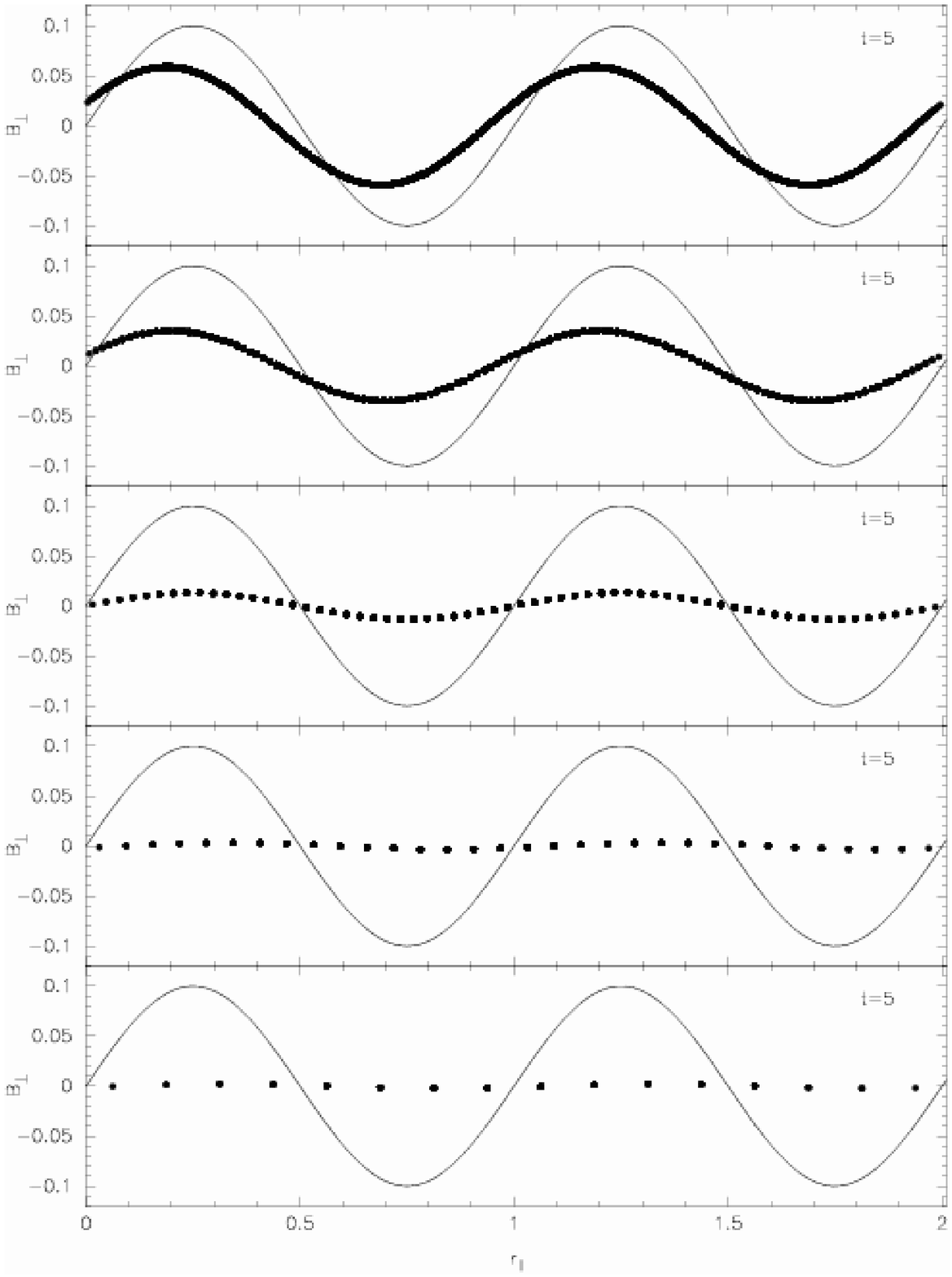,width=\textwidth}\end{turn}
\end{minipage}
\caption{Results of the circularly polarized Alfv\'en wave test at $t=5$ (corresponding to 5
wave periods). The plots show the perpendicular component of the magnetic field vector
$B_\perp = B_y \cos{\theta} - B_x \sin{\theta}$ for all of the particles,
projected against a vector parallel to the direction of wave propagation
$r_\parallel = x \cos{\theta} + y\sin{\theta}$ (where $\theta = 30^\circ$ in this
case). The SPMHD results are shown at five different resolutions which are, from
bottom to top, $8\times 16$, $16\times 32$, $32\times 64$, $64\times 128$ and
$128\times 256$. Initial conditions are indicated by the solid line. The numerical results
should match these initial conditions at the time shown. The left panel shows the results in the absence of dissipative terms and
demonstrates that the SPMHD algorithm contains very little intrinsic numerical dissipation even at low
resolutions, although there is a small phase error present even in the converged
higher resolution runs. The right hand panel shows the results applying the dissipative terms required in the shock tube problems uniformly (ie. in the absence of switches). In this case the wave amplitude is damped by the artificial resistivity term and exhibits somewhat slow convergence.}
\label{fig:malfvenresults}
\end{center}
\end{figure*}

 This test is described by \citet{toth00} where it is used to test a variety
of multidimensional MHD schemes in grid based codes. The test involves
a circularly polarized Alfv\'en wave propagating in a two
dimensional domain. The advantage of using a circularly (as opposed to
linearly) polarized wave is that it turns out to be an exact, non-linear solution to the
MHD equations, which means that the solution after one period should exactly
match the initial conditions, without the effects of nonlinear steepening (as
observed, for example, in the magnetosonic wave tests described in
paper~II). This also means that
the wave can be setup with a much larger amplitude than would be used for
purely linear waves.

 In \citet{toth00}, the wave is setup to propagate at an angle $\theta =
30^\circ$ with respect to the $x-$axis.  In SPH the orientation of the wave
vector with respect to the co-ordinates is not particularly important because
there is no spatial grid. However, we have retained the rotated configuration
as firstly it ensures that there are no spurious effects resulting from the
initial arrangement of the particles and secondly enables a fair comparison
with the results shown in \citet{toth00}. The particles are setup on a
hexagonal close packed lattice (ie. such that particles are equispaced) in a rectangular domain $0 < x < 1/\cos{\theta};
0 < y < 1/\sin{\theta}$. This positioning of the boundaries means that periodic boundary conditions can
be used, although some care is required to ensure the continuity of the lattice
across the boundaries. This is achieved by stretching the lattice slightly in
the $y-$direction to ensure that the boundaries lie at exactly half the spacing
of the rows in the lattice. The particle setup at the lowest resolution is shown in the left hand side of
Figure \ref{fig:malfvensetup}.

 The wave is setup with a unit wavelength along the direction of
propagation (ie. in this case along the line at an angle of $30^\circ$ with
respect to the x-axis). The initial conditions are $\rho = 1$, $P = 0.1$,
$v_\parallel = 0$, $B_\parallel = 1$, $v_\perp = B_\perp = 0.1\sin{(2\pi
r_\parallel)}$ and $v_z = B_z = 0.1 \cos{(2\pi r_\parallel)}$ with $\gamma =
5/3$ (where $r_\parallel = x \cos{\theta} + y\sin{\theta}$). The $x-$ and $y-$
components of the magnetic field are therefore given by $B_x = B_\parallel
\cos{\theta} - B_\perp \sin{\theta} $ and $B_y = B_\parallel \sin{\theta} +
B_\perp \cos{\theta}$ (and similarly for the velocity). Conversely,
$B_\parallel = B_y \sin{\theta} + B_x \cos{\theta}$ and $B_\perp = B_y
\cos{\theta} - B_x \sin{\theta}$. Note that this setup means that $\divB = 0$
holds as a combination of the $\partial B_x / \partial x$ and $\partial B_y /
\partial y$ terms, rather than both components being zero individually.
The vertical component of the magnetic field
after 5 periods is plotted as a rendered image in the right hand side of Figure
\ref{fig:malfvensetup}, showing the direction of wave propagation with respect
to the domain and the particle setup.

 We have peformed this test at five different resolutions: $8\times 16$,
 $16\times 32$, $32\times 64$, $64\times 128$ and $128\times 256$ particles. In each case the number of particles in the y-direction is
determined by the hexagonal lattice arrangement. The results are shown in Figure
\ref{fig:malfvenresults} after 5 wave periods (corresponding to $t=5$). The
plots show the perpendicular component of the magnetic
field $B_\perp$ plotted against $r_\parallel$ for all of the particles in the simulation, with the
results from the bottom to top panels shown in order of increasing
resolution. In each case the initial conditions are indicated by the solid line
which is identical to the exact solution at the time shown. 

 The left hand side
of Figure \ref{fig:malfvenresults} shows the results in the absence of
dissipative terms (that is with the artificial viscosity, resistivity and
thermal conductivity turned off). In this case the amplitude agrees very well
with the exact solution even at the lowest resolutions. This demonstrates that
SPH has a very low intrinsic numerical dissipation (compare for example with the
damping of the wave at lower resolutions in the plots shown in
\citealt{toth00}). However there is a small phase error which remains even in
the highest resolution run. This is similar to the phase error observed in the
one dimensional sound wave tests presented in \citet{price04} and in the
one dimensional magnetosonic waves tests in paper~I. In these cases
the phase error was found to be essentially removed by accounting for the variable smoothing length
terms (paper~II). The results shown in Figure
\ref{fig:malfvenresults} incorporate the variable smoothing length terms,
however in this case the phase error is not completely removed (although is
still an improvement over the results using simple averages of the smoothing lengths or
kernel gradients) unless a number of neighbours used is also increased in addition to the total number of particles.

 The right hand side of Figure \ref{fig:malfvenresults} shows the
results of this test using the dissipative terms as required in the shock tube
problems. In this case the wave is severely damped and convergence of the
amplitude towards the exact solution is quite slow. The damping is largely
caused by the uniform application of artificial resistivity (ie. using $\alpha_B
= 1$ everywhere) resulting in a somewhat large dissipation even in the absence
of shocks. Substantially improved results could be obtained using the resistivity
switch discussed in \S\ref{sec:avswitches}, however for the shock tube problems it
was found that use of such a switch could result in too little dissipation at
rotational discontinuities in the absence of a shear viscosity term. Nonetheless the results shown in Figure~\ref{fig:malfvenresults} suggest that some kind of resistivity switch would be very valuable in SPMHD calculations. Note that the divergence error remains very small $\left[(\divB)_{max}
\sim 10^{-3}\right]$ in all of the simulations shown even in the absence of any kind of divergence cleaning.
 
\subsection{2.5D shock tube}
\label{sec:25Dshock}
 The next two tests are simply two dimensional versions of the one dimensional shock tube
tests described in paper~I \citep[see also][]{price04} and demonstrate the effects of divergence
errors in the shock capturing scheme.  In two dimensions we setup the particles on a cubic lattice
in the $x-$direction in the domain $x = [-0.5-v_{x(L)} t_{max}, 0.5 -
v_{x(R)} t_{max}]$, where $v_{x(L)}$ and $v_{x(R)}$ are the initial
velocities assigned to the left and right states. This means that at the time $t_{max}$ the particles are contained
in the domain $x = [-0.5,0.5]$. The domain has a width of 4 particle spacings in
the $y-$direction for computational efficiency. Boundary conditions are implemented by fixing the
particle properties in two buffer regions at the edges of the $x-$domain, in
which particles are evolved with a fixed velocity but copy their properties
($\rho, P, \bB$) from
the nearest `active' particle. Periodic boundary conditions are used in the $y-$direction, implemented using
ghost particles. The exact position of the $y-$boundary is chosen to ensure
periodicity of the lattice arrangement, ie. at half the spacing of the initial
rows of particles in the y-direction. The initial shock is setup as a
discontinuity in the fluid quantities at $x=0$ to which no smoothing is applied. 

The first shock test is the adiabatic shock tube
problem involving seven different discontinuities given in paper~I. Strictly this is a
`$2\frac12$' dimensional problem since the transverse velocity and magnetic
field also have components in the $z-$direction.  Conditions to the left of the
discontinuity (the left state) are given by $(\rho,P,v_x,v_y,v_z,B_y,B_z) =
[1.08,0.95,1.2,0.01,0.5,3.6/(4\pi)^{1/2},2/(4\pi)^{1/2}]$ whilst to the right (the
right state) the conditions are $(\rho,P,v_x,v_y,v_z,B_y,B_z)=[1,1,0,0,0,4/(4\pi)^{1/2},2/(4\pi)^{1/2}]$ with $B_x =
2/(4\pi)^{1/2}$ everywhere and $\gamma=5/3$. The problem has been studied by
in one dimension by many authors \citep[e.g.][]{rj95,balsara98} and in two
dimensions by \citet{toth00} and \citet{dea02}.

\begin{figure*}
\begin{center}
\epsfig{file=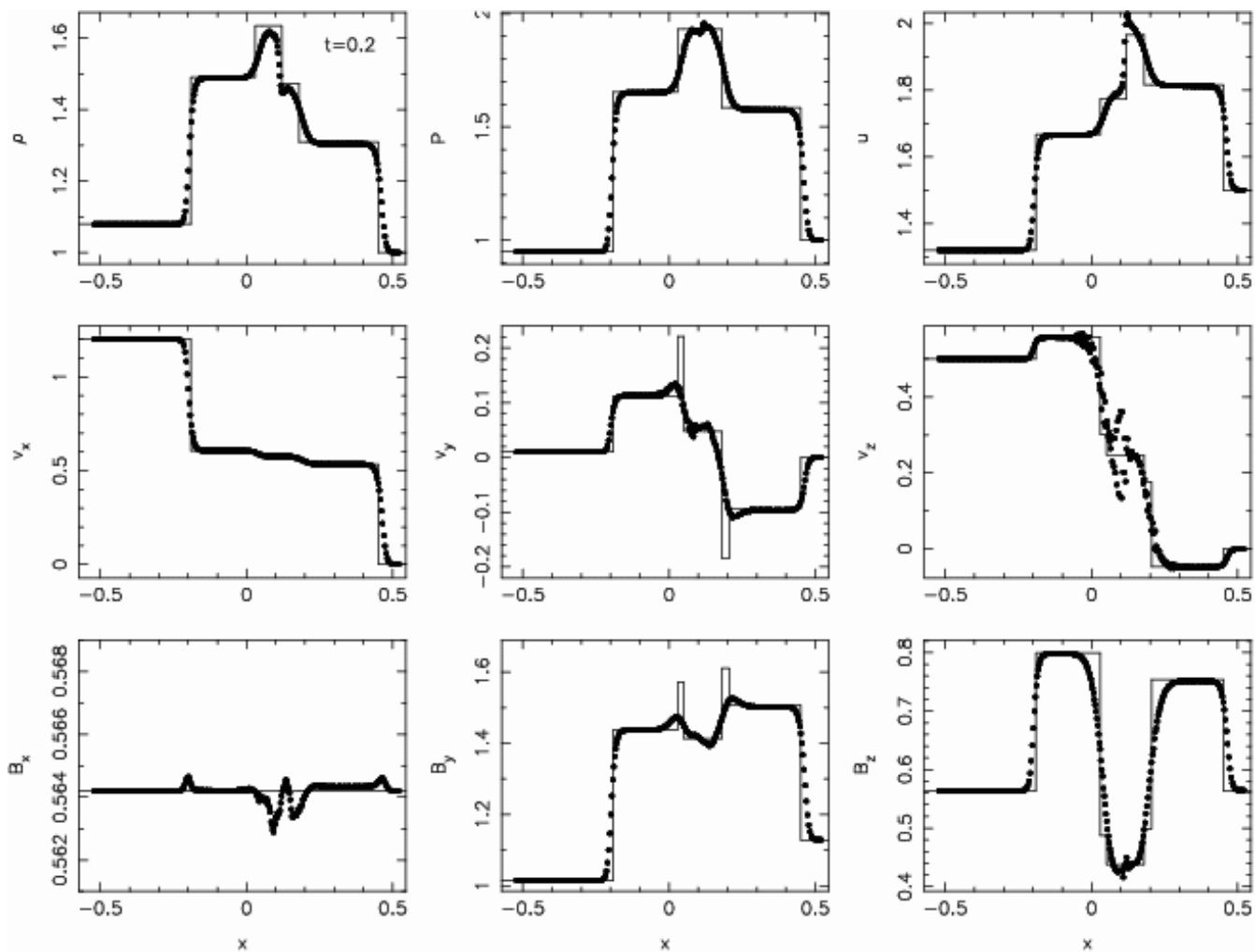,width=\textwidth}
\caption{Results of the 2.5D shock tube test using $310\times 4$ particles and an initial
smoothing length of $h = 1.2(m / \rho)^{1/2}$. In two dimensions at
this value of smoothing length small oscillations in the transverse velocity
components appear primarily as a result of the non-zero magnetic divergence. In this plot the usual artificial viscosity and resistivity terms have been applied uniformly (ie. not using switches). A small amount of artificial thermal conductivity has been applied using the switch.}
\label{fig:25Dshock}
\end{center}
\end{figure*}

\begin{figure*}
\begin{center}
\epsfig{file=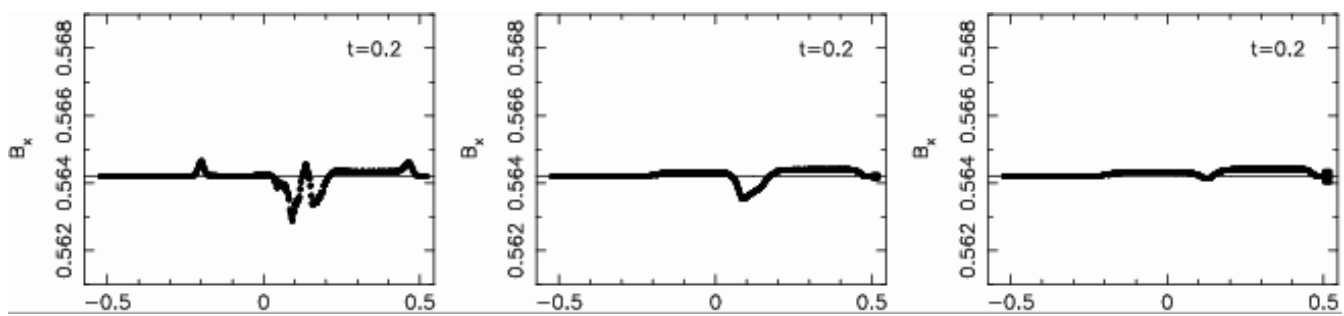,width=\textwidth}
\caption{The parallel component of the magnetic field in the 2.5D shock tube
problem using the usual dissipative terms (left),
using the total magnetic energy (centre) and using the total magnetic and
kinetic energies (right). Using the total magnetic energy in the dissipative
terms means that jumps in
the parallel field components are smoothed in addition to the jumps in
transverse field. Using the total kinetic energy smooths
jumps in the transverse (as well as parallel) velocity components, however this
explicitly adds an undesirable shear component to the artificial viscosity term.
Details of these formalisms are given in \S\ref{sec:mhdav}.}
\label{fig:25Dshock_Bx}
\end{center}
\end{figure*}

\begin{figure*}
\begin{center}
\epsfig{file=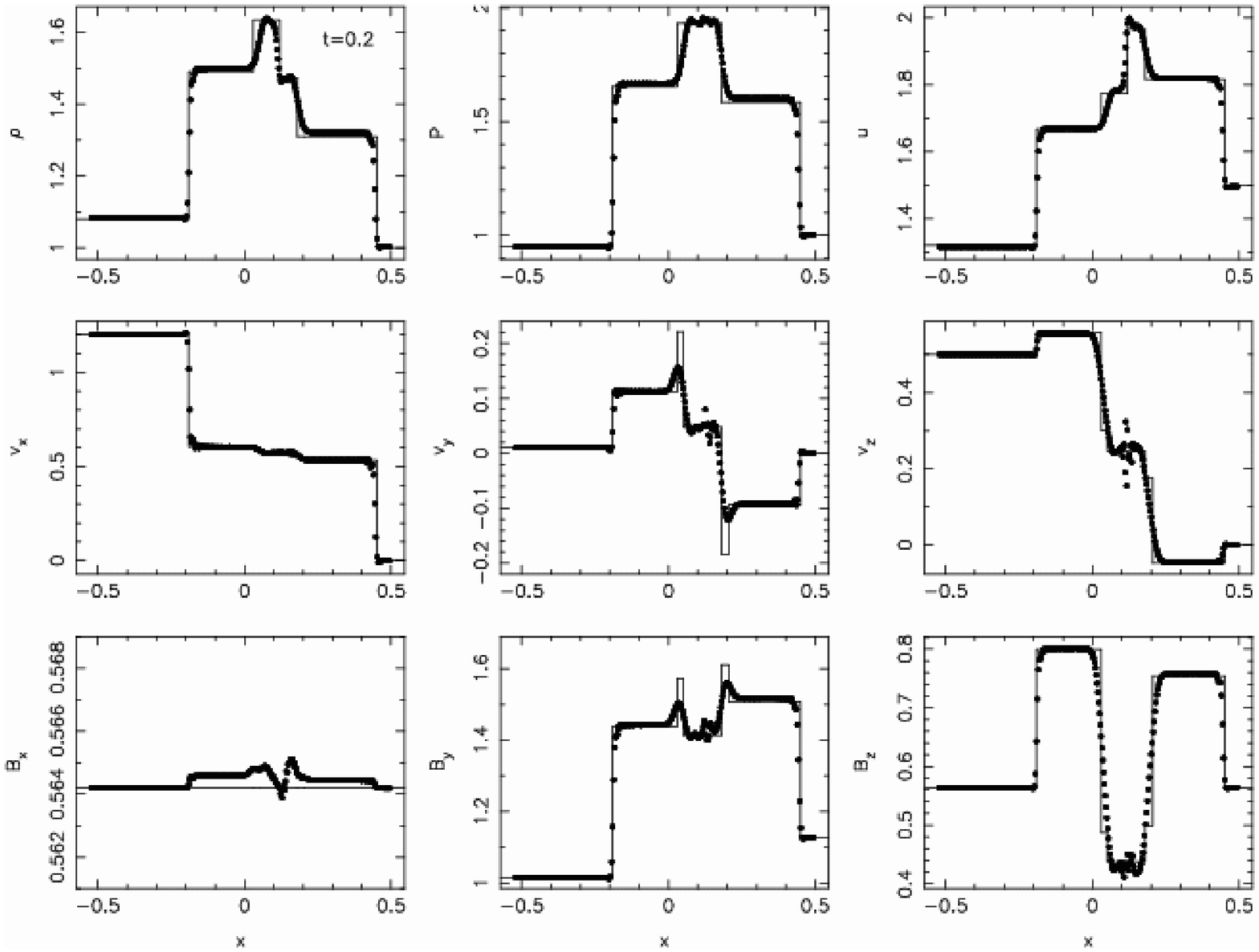,width=\textwidth}
\caption{Results of the 2.5D shock tube test using $310\times 4$ particles and an
initial smoothing length of $h = 1.5(m / \rho)^{1/2}$ and the total magnetic energy
in the artificial resistivity term but using the usual artificial viscosity
term (where in this case both have been applied using the dissipation switches). The results are a substantial improvement on those presented in Figure
\ref{fig:25Dshock} for a very modest increase in the number of neighbours.}
\label{fig:25Dshock_h1.5}
\end{center}
\end{figure*}

 The problem was computed using $310\times 4$
particles which corresponds to particles being uniformly spaced on a cubic lattice with
separation 0.004 (with a slightly larger spacing for $x > 0$ to give the density contrast), although results are similar
using a hexagonal close packed lattice arrangement. Note that the above figure refers to the number of particles in the domain $-0.5 < x
< 0.5$ at $t_{max} = 0.2$ and that the resolution in this domain is
correspondingly lower at earlier times due to the inflow boundary condition.
The resolution was chosen to be comparable to the resolutions used in
\citet{toth00}. The small density difference between
the left and right states was setup by changing the lattice spacing slightly in the $x-$direction.

 The solution using an initial smoothing length of $h = 1.2 (m / \rho)^{1/2}$ is
shown in Figure \ref{fig:25Dshock} at $t_{max}=0.2$ and may be
compared with the exact solution taken from \citet{rj95} (solid
line) and with the one dimensional SPMHD results shown in \citet{price04} and in paper~I.

 On the shock tests the most important physical aspects for a numerical algorithm are obtaining the correct physical states behind the shock fronts, since this represents the manner in which the gas is changed by the passage of the shock. Thus whilst the shock profiles shown in Figure~\ref{fig:25Dshock} are not as sharp as those shown at comparable resolution in, for example \citet{toth00}, the intermediate states are obtained correctly apart from some small oscillations observed in the transverse velocity components near the contact discontinuity and the very narrow spikes in the magnetic field and transverse velocities which are damped at this resolution by our use of artificial resistivity. 

 Whilst the damping due to artificial resistivity improves with resolution (and with the use of the resistivity switch -- see Figure~\ref{fig:25Dshock_h1.5}), the oscillations in transverse velocity are of more concern. It should be noted first of all that these oscillations are
quite small and do not appear to affect the dynamics significantly (mainly
because the jumps in the transverse velocity components are an order of magnitude less than
the jump in $v_x$). However, the oscillations appear to result from a combination of three factors: the unsmoothed initial
conditions, the fact that we do not explicitly apply any smoothing to the
transverse velocity components and the effects of the small jumps in the $x-$component of the
magnetic field.

 To remove these oscillations two approaches can be taken: The first approach is
to modify the artificial viscosity terms slightly in order to smooth the transverse velocity
profiles. The dissipative terms used in order to capture shocks were discussed at length
in paper~I and in this paper in \S\ref{sec:mhdav}. In the one dimensional case the
dissipation terms for MHD (comprising an artificial viscosity, artificial thermal conductivity and
artificial resistivity) were derived assuming that jumps would only occur in components of the
magnetic field transverse to the line joining the particles that jumps in
velocity would only occur parallel to this line. Neither of these assumptions
strictly hold in the shock tube problem shown in Figure \ref{fig:25Dshock} since
the transverse velocity components clearly jump and there is also a small jump
in the parallel field component due to the divergence errors.

 A reformulation of
the dissipative terms relaxing both of these assumptions was presented in
\S\ref{sec:mhdavtoten}, deriving the artificial viscosity and artificial
resistivity terms from jumps in the total kinetic and magnetic energies
respectively in the total energy equation. The effects of using these
formulations on the profile of the parallel component of the magnetic field are
shown in Figure \ref{fig:25Dshock_Bx}. From the centre panel we see that using the total magnetic energy
formulation for the artificial resistivity has clear advantages in preventing
oscillations in the parallel component of the field at shock fronts. Using the
total kinetic energy version of the artificial viscosity (in order to smooth out
jumps in the transverse velocity) effectively adds an
explicit shear component to the viscosity term.

 In \S\ref{sec:mhdavtoten} it was
noted that discontinuities in the transverse velocity components can only occur at corresponding jumps
in the magnetic field and therefore that such discontinuities are already
smoothed somewhat by the application of artificial resistivity there. For this
reason the total kinetic energy formalism is not strictly necessary provided that there is sufficient artificial resistivity present to smooth both the transverse field jumps and the transverse velocity jumps.
However, applying even a small amount of such a viscosity to the two dimensional problem is indeed found to remove the observed oscillations \citep{price04}. It is clear though that the use of this term is highly undesirable since applying an explicit shear viscosity will substantially increase the spurious transport of angular momentum caused by the artificial viscosity term.

 The second approach is to simply increase the number of neighbours slightly for each particle
to give a more accurate interpolation. The results using an initial smoothing
length of $h = 1.5 (m / \rho)^{1/2}$ are shown in Figure \ref{fig:25Dshock_h1.5}
using the total magnetic energy formulation of the artificial resistivity but retaining the
usual artificial viscosity formulation. In this case the jump in the parallel
field component is much lower and the oscillations in the transverse velocity
components do not appear, although there is a small glitch at the contact
discontinuity similar to that observed in the one dimensional case
\citep{price04}. The increase in smoothing length also means that the dissipative terms can be applied using the switches discussed in \S\ref{sec:avswitches}, resulting in a much lower dissipation rate away from the shocks than would be required for the $h=1.2 (m/\rho)^{1/2}$ case. 

Increasing the smoothing length from $h = 1.2 (m /
\rho)^{1/2}$ to $h = 1.5 (m / \rho)^{1/2}$ corresponds to an increase in the
number of neighbours from $\approx 20$ to $\approx 28$ on a uniform cubic
lattice in two dimensions. This quite a small increase in computational expense
for a substantial gain in accuracy (and stability). It therefore seems much more
desirable to increase the smoothing length slightly for multidimensional
problems rather than to explicit add a shear viscosity term.

 Finally, although this problem is not unstable to the clumping instability (and indeed no clumping is observed) we have also
investigated the effects of various instability correction methods on the shock
profile. In particular use of the anticlumping term (paper~I) was
found to produce additional noise in the shock profile. Using either the Morris
formalism for the anisotropic force (\S\ref{sec:otherposs}) or subtracting the
constant component of the magnetic field (\S\ref{sec:subtractBconst}) both give results
very similar to those shown in Figures \ref{fig:25Dshock}-\ref{fig:25Dshock_h1.5}.
Applying the hyperbolic/parabolic divergence cleaning to this problem gives a small reduction in the divergence error but otherwise has no significant effect on the shock profiles.

\subsection{Two dimensional shock tube}
\label{sec:shock2D}
 The second shock tube test is used by both \citet{toth00} and
\citet{dea02} in two dimensions to compare the results of various divergence
cleaning schemes, although the one dimensional version of this test has been
used by many authors \citep[e.g][]{dw94,rj95}. The results of the one dimensional
test using the SPMHD algorithm are presented in \citet{price04}. Although this is a purely two dimensional test we
present it after the 2.5D shock tube since it presents a much more challenging
problem with regards to the non-zero divergence of the magnetic field due to the
stronger shocks.

\begin{figure*}
\begin{center}
\epsfig{file=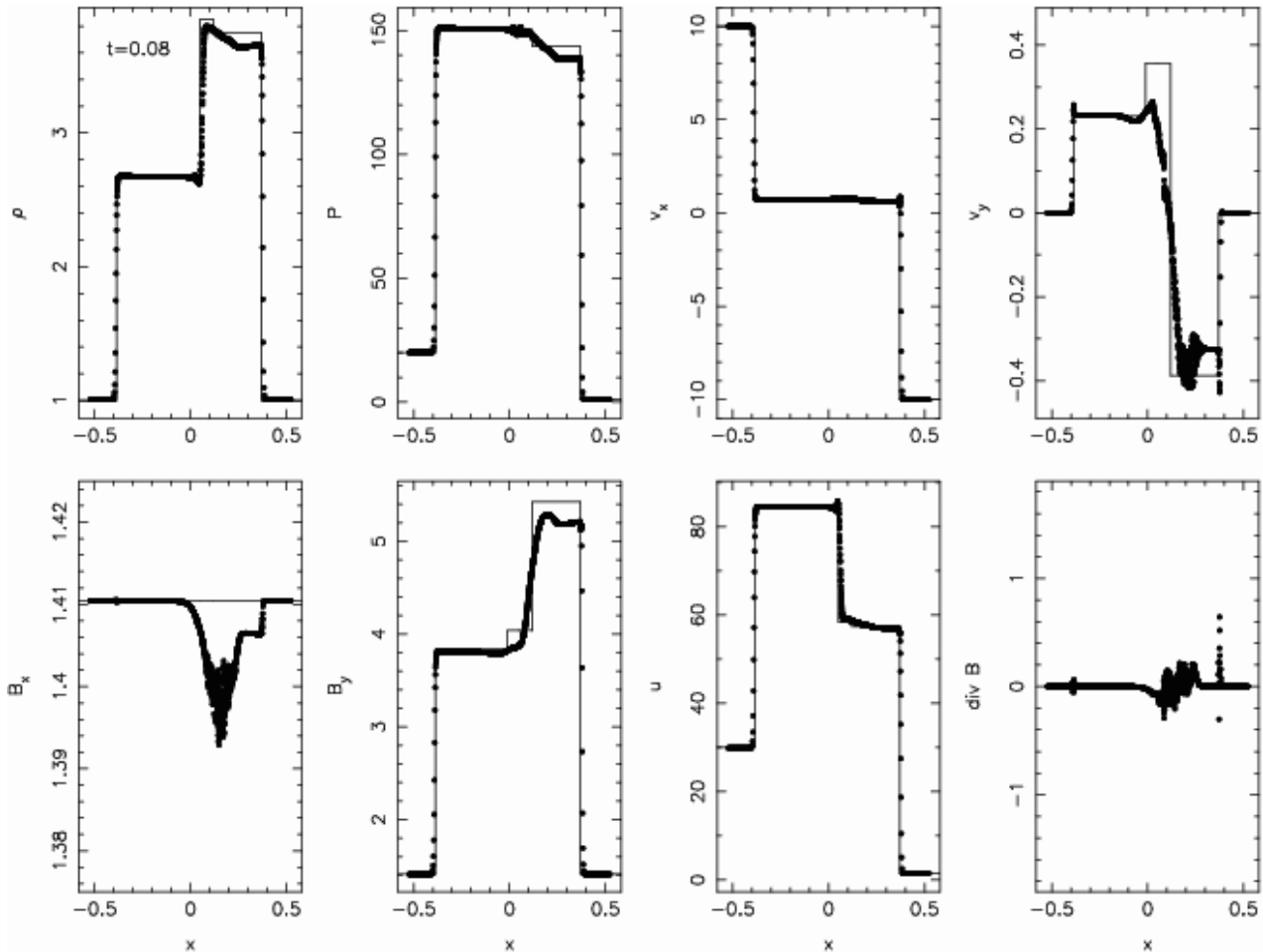,width=0.95\textwidth}
\caption{Results of the two dimensional shock tube test at $t=0.08$ using $h =
1.5 (m /\rho)^{1/2}$ and the shear viscosity term. The results may be compared
with the exact solution given by the solid line. In this stronger shock tube
problem the jumps in the parallel field can cause significant oscillations in
the transverse velocity components due to the non-zero divergence terms.}
\label{fig:tothshock2D}
\end{center}
\end{figure*}

 The particle setup is as described in the previous section, except that the
initial left state is given by $(\rho,P,v_x,v_y,B_y) =
[1,20,10,0,5/(4\pi)^{1/2}]$ and the right state is
$(\rho,P,v_x,v_y,B_y) = [1,1,-10,0,5/(4\pi)^{1/2}]$ with
$B_x = 5.0/(4\pi)^{1/2}$ and $\gamma = 5/3$. The boundaries are correspondingly
adjusted in the $x-$direction to allow the particles to fill the domain $-0.5 < x
< 0.5$ at $t_{max} = 0.08$. Particles are arranged initially on a hexagonal lattice
with particle spacing 0.004, giving 660 particles in the $x-$direction and a
total particle number of $660\times4 = 2640$. As in the
previous test, the results using an initial smoothing length of $h = 1.2 (m /
\rho)^{1/2}$ exhibit significant oscillations in the transverse velocity
($v_y$). In this case the oscillations are substantially worse because the jump
in the parallel field component is much larger. Hence
we have performed this test using $h = 1.5 (m /\rho)^{1/2}$. However, even in
this case the oscillations remain present and so we have also added the
shear viscosity term, using (\ref{eq:vdissfull}) with $\alpha = 1$ everywhere
(that is, not using the viscosity switch). The results using these settings are shown in Figure \ref{fig:tothshock2D} and
may be compared with the exact solution taken from \citet{dw94} (solid line) and
with the one dimensional results given in \citet{price04}. All particles are shown projected along the $x-$direction. Even in this case some oscillations are visible in the $v_y$ profile, corresponding
exactly with a spike in $\divB$. In the $h = 1.2 (m /\rho)^{1/2}$ case this
spike is much larger [$(\divB)_{max} \sim 40$], causing significantly more
disruption to the velocity profile. Thus despite the various tweaks we have
attempted for this test, the oscillations appear to be primarily caused by the divergence
errors generated at the shocks. More importantly a slight offset in the intermediate states in the $v_{y}$, $B_{y}$, $\rho$ and $P$ profiles is present. Investigation of this effect suggests that this is caused by a combination of the divergence error and the shear viscosity term. Without the shear viscosity term the intermediate states \emph{are} obtained correctly, although with substantially more oscillations in the $v_{y}$ profile.

 The effects of increasing the number of neighbours and changing the strength of
the dissipation terms may be summarised as follows: Increasing the number of
neigbours reduces the jumps in the parallel field component (for example with   
$h= 1.2(m /\rho)^{1/2}$ the jump is given by $\Delta B_x = [B_{x(max)} -
B_{x(min)}]/B_{x0} \approx 18\%$ whilst for $h= 1.5(m /\rho)^{1/2}$ we have $\Delta B_x
\approx 1\%$ and for $h=2.4(m /\rho)^{1/2}$ this reduces further still to $\Delta B_x
\approx 0.15\%$). On the other hand, adding dissipation at the jumps in parallel
field means that although such jumps may be present, the discontinuities (causing strong
divergence errors) are smoothed. The effect of adding the shear viscosity term
is to increase the dissipation at these discontinuities, thus reducing to some
extent the associated spike in the magnetic divergence.

 In \citet{toth00} the results of this test were presented using the source
term approach of \citet{pea99} (discussed in \S\ref{sec:monopoles}), showing similar jumps in
the parallel magnetic field component which were unchanged even in the converged
numerical results. The fact that the jumps in parallel field
reduce with an increasing number of neighbours indicates that the SPMHD
algorithm converges to the exact solution in the limit of $h\to \infty$ and
$N\to \infty$ where $N$ is the number of particles. \citet{toth00} attributes
the errors in the parallel field components in the Powell method to the
non-conservative source terms in the induction equation. We have also performed
this simulation using the `conservative' induction equation (\ref{eq:indconssph}), however we find
that the jumps in $B_x$ are not changed significantly by
including the $\bv \divB$ term (although contain substantially more numerical
noise).
\begin{figure}
\begin{center}
\epsfig{file=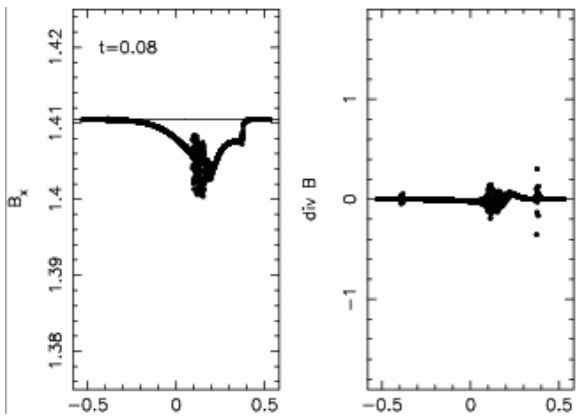,width=0.9\columnwidth}
\caption{Parallel magnetic field (left) and the divergence error (right) in the two dimensional shock tube test at $t=0.08$ computed as in Figure \ref{fig:tothshock2D} but using the hyperbolic/parabolic divergence cleaning (\S\ref{sec:hyperbolic}). The exact solution is given by the solid line. The hyperbolic divergence cleaning does not have a large effect on this problem since the divergence errors are
propagated at the fastest wave speed which is similar to the rate at which they
are generated in the shocks.}
\label{fig:tothshock2D_hyp}
\end{center}
\end{figure}

\begin{figure*}
\begin{center}
\epsfig{file=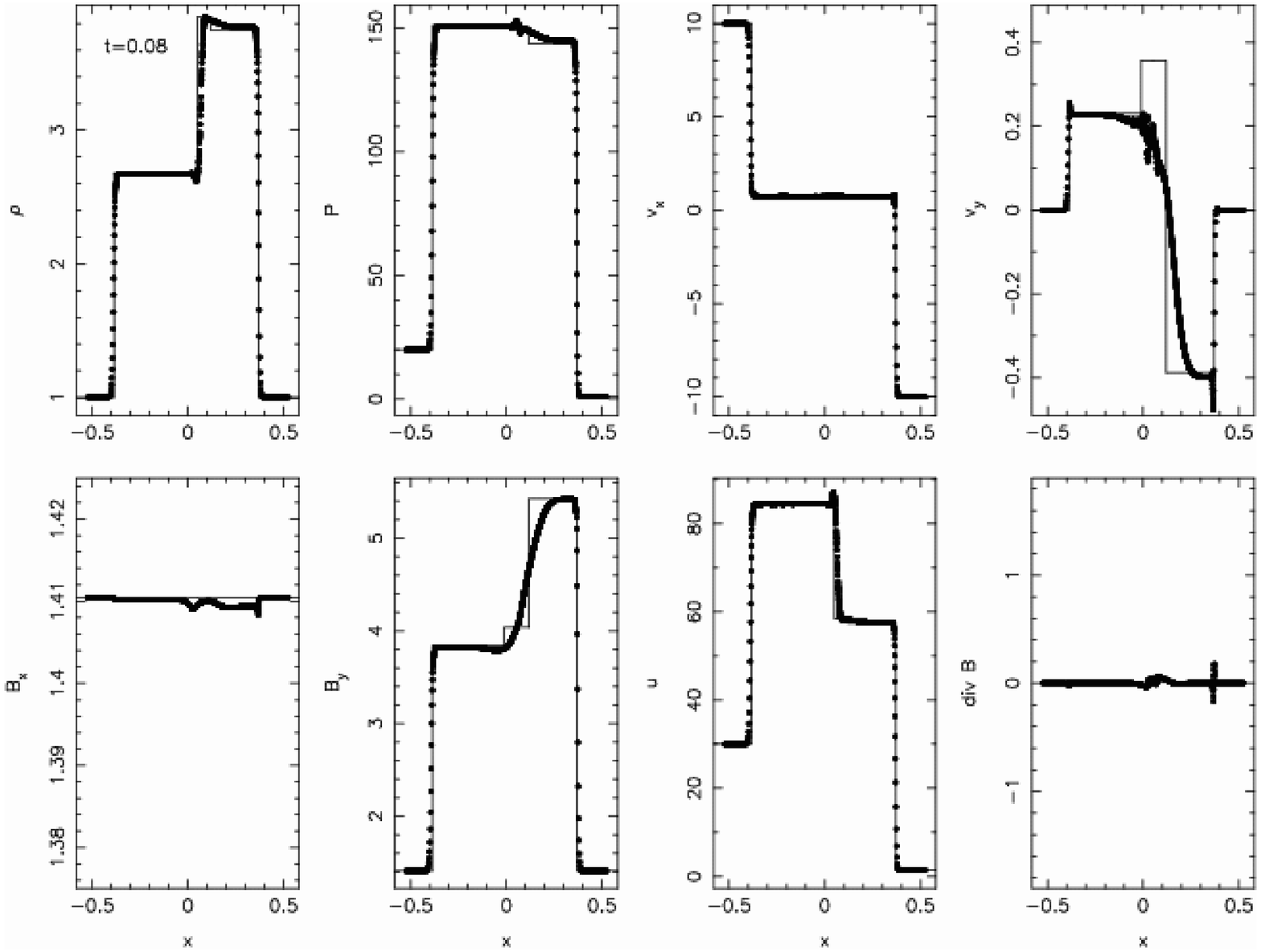,width=0.95\textwidth}
\caption{Results of the two dimensional shock tube test at $t=0.08$ computed as
in Figure \ref{fig:tothshock2D} but using $h = 2.4 (m/\rho)^{1/2}$. The exact solution is given by the solid line. Increasing the neighbour number significantly decreases the divergence error.}
\label{fig:tothshock2D_hbig}
\end{center}
\end{figure*}

 The shock tube tests presented above have been computed without using any form
of divergence cleaning (other than the consistent formulation of the MHD
equations in the presence of magnetic monopoles discussed in
\S\ref{sec:monopoles}). Thus a way of eliminating both the jumps in parallel field and the
resulting oscillations in the transverse velocity components is to clean up the
divergence error. Using the hyperbolic/parabolic cleaning discussed in
\S\ref{sec:hyperbolic} is not particularly effective for this problem, since the
divergence errors are propagated away from their source at the fastest wave speed which is similar to
the rate at which they are created by the shocks. Thus the diffusion introduced
by the parabolic term does not have time to eliminate the divergence error
before oscillations in the velocity components are produced. This is illustrated
in Figure \ref{fig:tothshock2D_hyp} which shows the parallel field component and the divergence error after using this type of cleaning with $\sigma = 0.4$ in the
parabolic term (c.f. \S\ref{sec:Bxpeaktest}). The divergence
errors are reduced by a factor of $\approx 2$ compared to the results shown in
Figure \ref{fig:tothshock2D}. In order to eliminate the divergence errors from
problems such as this one where divergence errors are created rapidly it seems necessary to invoke some kind of sub-timestep cleaning (such as a projection method). The implementation of such methods are complicated in this simple test problem by the use of periodic boundary conditions. Alternatively the number of neighbours can be increased further. To demonstrate this we present a simulation at double the usual neighbour number, that is using $h = 2.4 (m/\rho)^{1/2}$. The results are shown at $t=0.08$ in Figure~\ref{fig:tothshock2D_hbig} and show a reduction in the divergence error by a factor of $\sim 10$ compared to the results shown in Figure~\ref{fig:tothshock2D}. This suggests that using a larger neighbour number may be crucial in three-dimensional SPMHD simulations.

\subsection{Rotor}
\label{sec:mhdrotor}
\begin{figure*}
\begin{center}
\epsfig{file=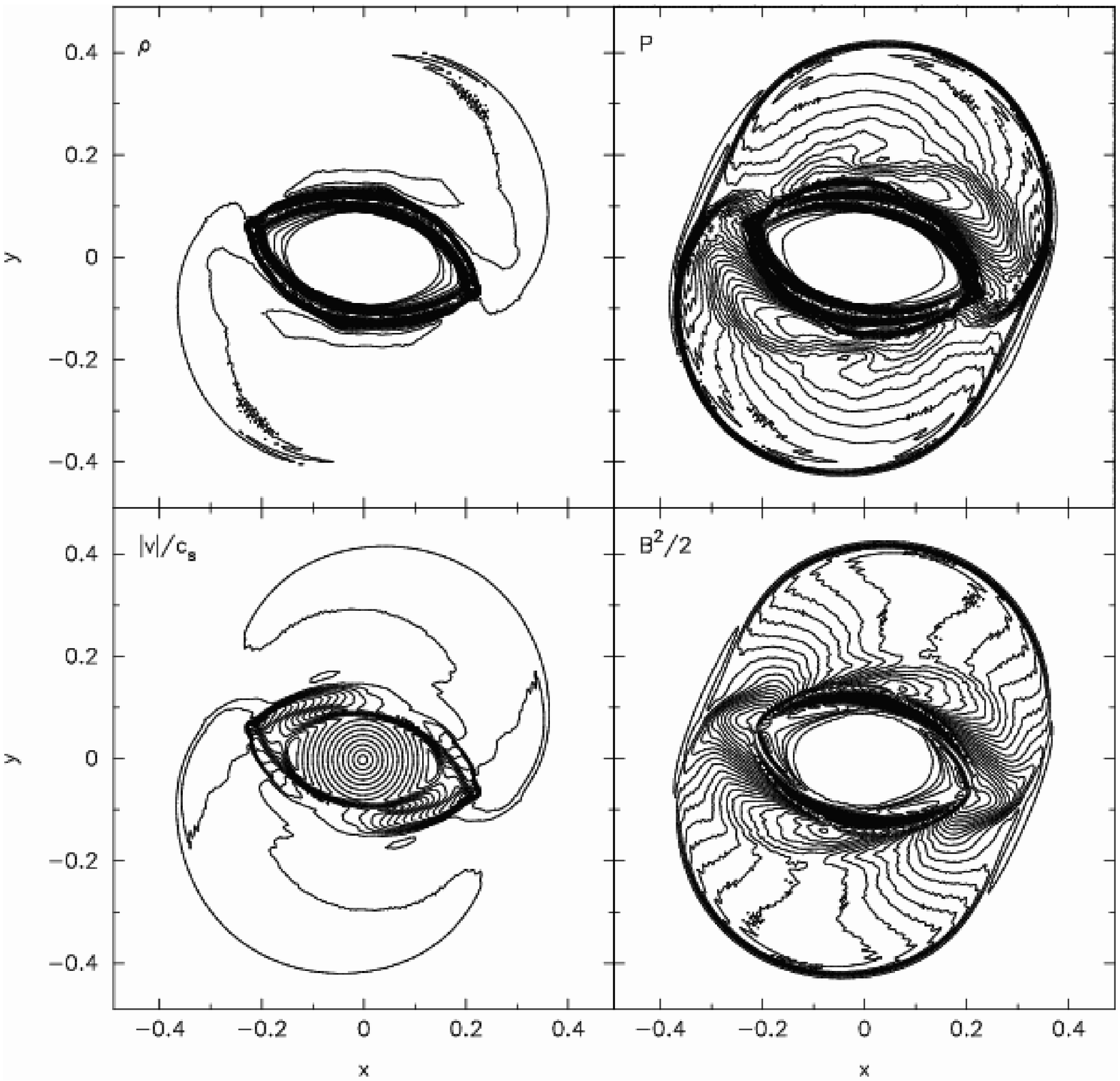,height=0.8\textwidth}
\caption{The density, pressure, mach number and magnetic pressure at $t=0.15$ in the MHD rotor problem using 58,786 particles (roughly $200 \times 200$). All plots show 30 contours spaced between the limits given in \citet{toth00}, that is $0.483 < \rho < 12.95$, $0.0202 < P < 2.008$, $0 < \vert v\vert /c_{s}< 1.09$, $0 < \frac12 B^{2} < 2.642$ in order to make a direct comparison.}
\label{fig:mblast2D}
\end{center}
\end{figure*}

 The next test is taken from \citet{toth00} and consists of a spinning, dense disc embedded in an ambient background medium containing a uniform magnetic field. The material initially contained within the disc is flung into the surrounding medium by the centrifugal forces, but is constrained into an oblate shape by the magnetic field. The computational domain is given by $-0.5 < x, y < 0.5$ with uniform thermal pressure $P=1$ and an adiabatic index of $\gamma=1.4$. A constant, uniform magnetic field is setup in the $x-$direction with strength $B_{x}=5/\sqrt{4\pi}$. The dense disc is setup with $\rho=10$ and a rotation velocity given by $v_{x} = 2(y-0.5)/r_{0}$, $v_{y} = 2(x-0.5)/r_{0}$ for $r < r_{0}$ where in this case $r_{0} = 0.1$. The ambient medium is at rest with $\rho = 1.0$. Note that this choice of initial conditions corresponds to the `first rotor problem' in \citet{toth00}.
 
  The density contrast between the disc and the background medium can be setup in SPH using either variable particle masses and therefore a fixed initial separation or equal mass particles and a variable particle distribution. We have experimented with both methods. In the variable particle mass case the large density contrast results in some spurious effects from the higher mass particles `mixing' into the low particle mass medium and we therefore prefer the equal mass particle approach. We achieve this setup by setting up the initial disc with a dense concentration of particles setup on a regular, hexagonal close-packed lattice trimmed to $r < r_{0}$. The surrounding medium is then placed using a second close packed lattice with a correspondingly larger inter-particle separation with the region $r < r_{0}$ excluded. This setup means that we do not apply a taper function to the density, pressure or velocity profiles as in \citet{toth00}. However the density profile is naturally tapered by the iterative calculation of the smoothing lengths and densities of the particles across the interface (\S\ref{sec:spmhd}). To ensure numerical pressure equilibrium we setup the thermal energy of the particles using $u_{0} = P_{0}/[(\gamma - 1)\rho_{0}]$ after the initial density has been calculated by direct summation, rather than using the analytic density step. Despite this there are some initial transients but these do not appear to affect the subsequent evolution substantially.
   
  The problem has been calculated using a background medium with $200$ particles in the $x-$direction. The hexagonal lattice arrangement means that this corresponds to $200\times230$ particles in the surrounding medium, from which the central disc region is removed, leaving 44,332 particles in the background medium. The dense concentration of particles in the disc contains a further 14,454 particles, resulting in a total of 58,786 particles. Artificial viscosity and resistivity have been applied using the switches, with artificial thermal conductivity turned off. No divergence cleaning has been applied.
  
  The results at this resolution using $h=1.2(m/\rho)^\frac12$ are plotted in Figure~\ref{fig:mblast2D} and may be directly compared with the high resolution results shown in \citet{toth00}. The density resolution in the SPMHD solution is slightly better than even the $400\times 400$ grid based solution shown in \citet{toth00}, giving a maximum density of $\rho_{max} = 15.54$ at $t=0.15$ as opposed to $\rho_{max} = 12.95$ in the grid solution, although the minimum density at this time is $\rho_{min} = 0.74$ in the SPMHD solution as opposed to $\rho_{min}=0.483$ in \citet{toth00}. The SPMHD result using 400 particles in the $x-$direction (giving a total of 235,574 particles) resolves a density range of $\rho_{max}=17.76$ and $\rho_{min} = 0.58$. The maximum field strength is a little lower in the SPMHD calculations, with $(\frac12 B^{2})_{max} = 2.3$ (or $2.45$ at the higher resolution), as opposed to $(\frac12 B^{2})_{max} = 2.64$ in the $400\times 400$ grid solution. This is due to our use of artificial resistivity for shock capturing. There are some small effects at low densities in the SPMHD solution due to the particle distribution. These effects decrease both with particle number and also as the number of neighbours is increased. The divergence constraint is maintained reasonably well in this problem -- for example $95\%$ of the particles have $h\vert\divB\vert/\vert{\bf B}\vert < 0.01$ in the $200\times 200$ particle simulation, which increases to $98\%$ using $400\times 400$ particles and decreases to $87\%$ using $100\times 100$ particles. 

\subsection{Orszag-Tang vortex}
\label{sec:orstang}
 The final test is the compressible Orszag-Tang vortex problem which was
first investigated by \citet{ot79} in order to study incompressible
MHD turbulence. The problem was later extended to the compressible case by
\citet{dp89} and \citet{pd91}. More recently it has been widely used as a test problem
for multidimensional MHD algorithms \citep[e.g.][]{rea95,balsara98,dw98,ld00,toth00}.

\begin{figure*}
\begin{center}
\epsfig{file=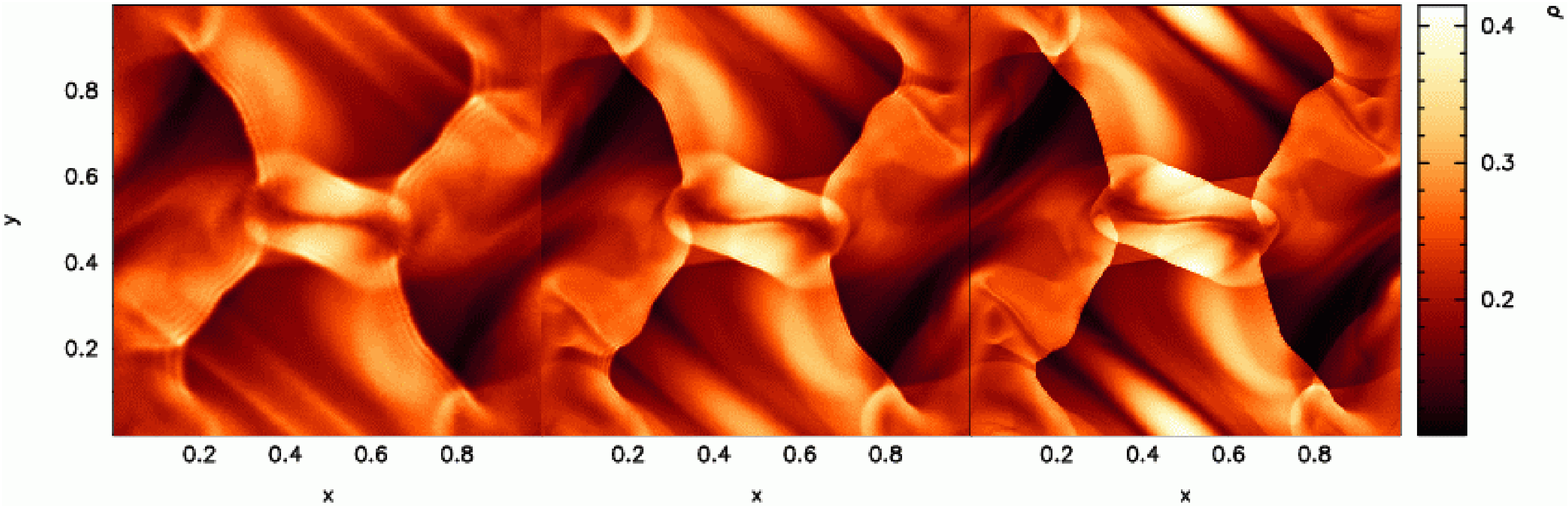,width=1.1\textwidth}
\vspace{2mm}
\epsfig{file=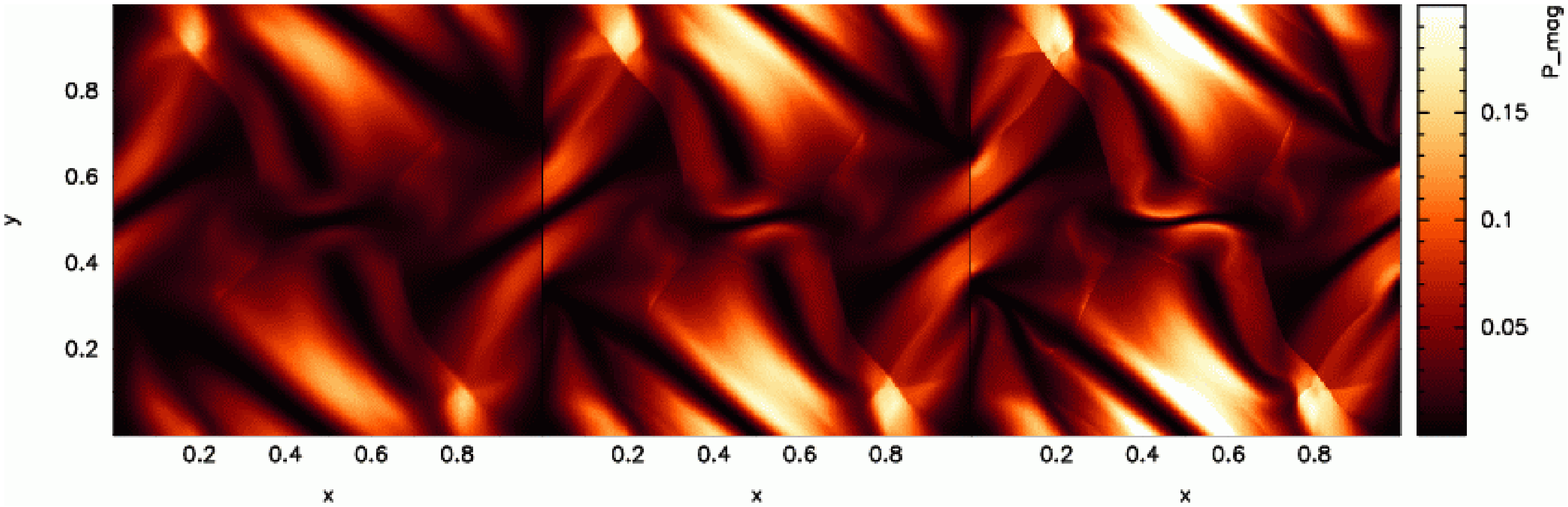,width=1.1\textwidth}
\caption{Results of the two dimensional Orszag-Tang vortex test, showing
the density (top) and magnetic pressure (bottom) distributions at $t=0.5$ for resolutions of $128\times 146$ (left), $256\times 294$(centre) and $512\times 590$(right) particles. The particles are
initially arranged on an isotropic hexagonal lattice with periodic boundary conditions. 
The initial velocity field is a large vortex $\bv = [-\sin{(2\pi y)},\sin{(2\pi x)}]$
whilst the magnetic field has a doubly periodic geometry $\bB = B_0[-\sin{(2\pi y)},\sin{(4\pi x)}]$.
The SPMHD results at higher resolutions are in excellent agreement with those
presented in (e.g.) \citet{dw98} and \citet{toth00}.}
\label{fig:orstang}
\end{center}
\end{figure*}

 The setup consists of an
initially uniform density, periodic $1 \times 1$ box given an initial velocity
perturbation $\bv = v_0[-\sin{(2\pi y)},\sin{(2\pi x)}]$ where $v_0 = 1$. The magnetic field is given
a doubly periodic geometry $\bB = B_0[-\sin{(2\pi y)},\sin{(4\pi x)}]$ where $B_0 =
1/\sqrt{4\pi}$. The flow
has an initial average Mach number of unity, a ratio of magnetic to thermal
pressure of $10/3$ and we use $\gamma = 5/3$. The initial gas state is therefore $P = 5/3 B_0^2 =
5/(12\pi)$ and $\rho = \gamma P/v_0 = 25/(36\pi)$.
Note that the choice of length and time scales differs slightly between various
implementations in the literature. The setup used above follows that of \citet{rea95} and
\citet{ld00}.

 The particles are arranged initially on a uniform hexagonal close packed lattice. This
distribution means that the particle are isotropically arranged and is the distribution towards which other
arrangements naturally settle. However, results are similar using a cubic lattice arrangement. The
simulation is performed at three different resolutions: $128\times 146$, $256\times 294$ and $512\times 590$ particles (where the number of particles in
the $y-$direction is determined by the isotropic lattice arrangement). The periodic boundary conditions
are implemented using ghost particles. These resolutions are similar to the resolutions used in \citet{dw98} (although in SPH the resolution is
concentrated preferentially towards regions of high density). The dissipation terms are
applied using the artificial viscosity and resistivity switches but leaving the artificial thermal conductivity turned off in order to increase the density resolution. The wall heating effects which the artificial thermal
conductivity prevents are discussed in \citet{price04} and are in general quite minor. No shear viscosity term has been used. Simulations of this problem which have been run with or without the
variable smoothing length terms, using the Morris formalism for the magnetic force
(\S\ref{sec:otherposs}), evolving either $\bB$ or $\bB/\rho$ and either the
thermal or total energy show essentially no difference in the numerical results.

\begin{figure}
\begin{center}
\begin{turn}{270}\epsfig{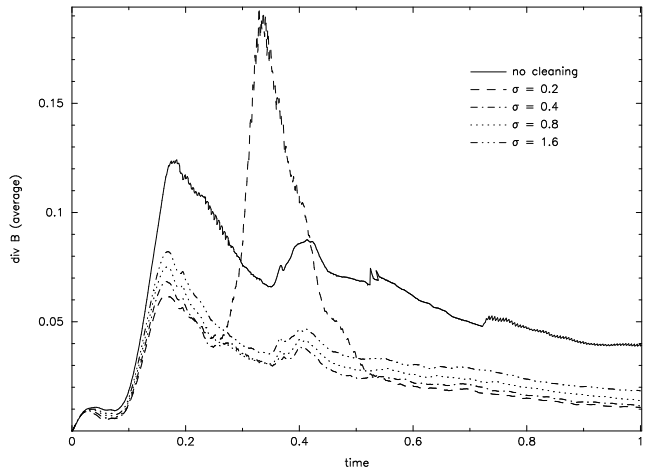}\end{turn}
\caption{Effect of the hyperbolic/parabolic cleaning on the evolution of the average magnetic divergence in the two dimensional Orszag-Tang vortex problem, varying the parameter $\sigma$. With $\sigma$ too low the cleaning can cause increases in $\divB$ (dashed line) over simulations with no divergence cleaning (solid line). The optimal cleaning is obtained with $\sigma \sim 0.4-0.8$ (dot-dashed, dotted lines). However, the reduction in the divergence obtained using the hyperbolic cleaning is fairly small.
The single biggest factor determining the magnitude of the divergence error
is the number of neighbours. The results shown are for a smoothing length of $h = 1.2 (m/\rho)^{1/2}$, although the errors decrease as the number of neighbours is increased.}
\label{fig:orstang_divb}
\end{center}
\end{figure}

 The results of the density evolution are shown in Figure \ref{fig:orstang} at $t=0.5$.
At this time four main shock fronts are visible which have interacted in the central regions after having
crossed the periodic domain. The SPMHD results, particularly  are in good agreement with those
presented in (e.g.) \citet{dw94,dw98} and \citet{toth00}. In the lowest resolution run, the central regions appear to be slightly better resolved than in the $128\times 128$ fixed-grid simulation of \citet{dw98},
although the lower density regions are correspondingly less well resolved. At this resolution the SPMHD solution shows some small residual effects due to the distortion of the initial regular particle
arrangement, noticable as small ripples behind the shock fronts in Figure
\ref{fig:orstang} and a slightly mottled appearance in the low density regions. This is particularly evident in Figure \ref{fig:orstang} since we have used a smoothing length of $h = 1.2 (m/\rho)^{1/2}$. In the lowest resolution run the density maxima visible in the higher resolution runs at the top and bottom of the domain are largely washed out. This is a result of the artificial resisitivity term used for shock capturing which dissipates energy in these regions due to the strong current gradient.

 The evolution of the average of the magnetic divergence is shown in Figure
\ref{fig:orstang_divb} for several runs using the hyperbolic divergence cleaning. The results
using the hyperbolic/parabolic cleaning with $\sigma = 0.2$ (dashed line) can in fact increase the divergence error over the results with no divergence cleaning
(solid line). This is because in the absence of sufficient diffusion the hyperbolic term can spread the divergence errors such that the resultant `divergence waves' can constructively interfere with each other, leading to increased errors. The optimal cleaning is obtained with $\sigma \sim 0.4-0.8$ (dot-dashed, dotted lines), although the reduction in the divergence error given by the hyperbolic cleaning is comparatively small.
In fact, as in the previous tests, the single biggest factor which determines the magnitude of the divergence error is the number of
neighbouring particles. For example in a simulation using $h = 1.5 (m/\rho)^{1/2}$ the
divergence errors are approximately half those shown in Figure \ref{fig:orstang_divb}.

\section{Discussion}
\label{sec:summary}
 In this paper multidimensional aspects of the SPMHD algorithm have been discussed. In
particular several methods for maintaining the divergence-free constraint in an SPH
context have been presented. Firstly the source term approach
of \citet{pea99} was outlined and contrasted with the consistent formulation of
the MHD (and SPMHD) equations derived in paper~II. The major difference between
the two approaches is that our approach retains the conservation of momentum and
energy whereas the \citeauthor{pea99} approach does not. The conservation
properties of the induction equation were also discussed, in which it was
highlighted that using a `non-conservative' induction equation means that the
surface integral of the magnetic flux is conserved, rather than the volume
integral. The effect of using the consistent formulation of the MHD equations in
the presence of magnetic monopoles (which conserves the surface integral of the
flux) is that divergence errors are advected without change by the flow (illustrated in
Figure \ref{fig:divbpeak1}). 

 Projection methods for maintaining a divergence free field were discussed in an
SPH context in \S\ref{sec:projection}. In particular it was noted that using the
Green's function solution to the Poisson equation (as is often used for
self-gravity in SPH) provides only an approximate projection. The results using
this type of projection on a problem where an initial magnetic divergence was
introduced into the simulation were very good
(\S\ref{sec:Bxpeaktest}), but were found to degrade as the wavelength of the divergence error approached the resolution length. A projection method based on Biot-Savart's law was also discussed and found to give excellent results even for wavelengths approaching the smoothing length. The implementation of either of these projection schemes for the test problems considered in this paper was complicated by the periodic boundary conditions used, leaving a need for further testing of these methods on three dimensional problems. In particular the Biot-Savart projection method suggests a promising divergence cleaning method in three dimensions. This would however require implementation in a tree-code which is beyond the scope of this paper.

An alternative approach to divergence cleaning suggested recently by
\citet{dea02} was discussed in \S\ref{sec:hyperbolic}. The method involves
adding an additional constraint equation which is coupled to the induction
equation for the magnetic field. Chosen appropriately, the effect of this
equation is to cause the divergence errors to be propagated in a wave-like
manner away from their source (Figure \ref{fig:divbpeak1}). Adding a small diffusive term means that the
divergence errors are also rapidly reduced to zero. This method is extremely
simple to implement and is computationally very inexpensive. The disadvantage is
that the error propagation is limited by the timestep condition
and, although much faster than using diffusion alone to reduce the divergence, for some problems (for example the shock tube tests given in
\S\ref{sec:25Dshock} and \S\ref{sec:shock2D}) the cleaning is still
not fast enough. However, this method is some improvement over not
using any form of divergence cleaning at a negligible additional computational
cost.

 The SPMHD algorithm was also tested against a variety of multidimensional
test problems. A non-linear circularly polarized Alfv\'en wave was studied in
\S\ref{sec:alfven}. This test showed that SPMHD has a very low intrinsic
numerical dissipation compared to grid based codes, although this property is
destroyed by the addition of explicitly dissipative terms for shock-capturing which can
cause quite slow convergence on problems where the physical dissipation
timescale is of critical importance.

 Two of the shock tube problems examined previously in
one dimensional simulations \citep[paper~I][]{price04} were examined in two dimensions in \S\ref{sec:25Dshock}
and \ref{sec:shock2D}. For these problems jumps in the component of the magnetic
field parallel to the shock front (causing divergence errors) were found to
result in oscillations in the transverse velocity profiles. The jumps in the
parallel field component were found to decrease as the number of neighbours for
each particle was increased. The corresponding divergence errors produced by these jumps could be
reduced by using a form of the dissipative terms derived in
\S\ref{sec:mhdavtoten} using the total jump in magnetic and kinetic energies.
Modifying the artificial viscosity term in this manner results in the addition
of an explicit shear viscosity component. It is therefore somewhat
undesirable to do so since this can result in excess spurious angular momentum
transport elsewhere. A better approach would be to use divergence cleaning to
prevent these errors from occuring. However, the hyperbolic cleaning was not
found to be particularly effective for this problem because of the restriction
to the fastest wave speed and implementation of the
projection method is complicated by the periodic boundary conditions. These
difficulties are not, however, insurmountable. The single biggest factor in determining
the magnitude of the divergence errors in the shock tube tests was found to be the size
of the smoothing region (ie. the number of contributing neighbours). It therefore
seems advantageous to use a slightly larger number of neighbours for MHD
problems (typically $h \gtrsim 1.5 (m/\rho)^{1/\nu}$ where $\nu$ is the number of spatial
dimensions) than might otherwise be used for hydrodynamics.

 An MHD rotor problem was examined in \S\ref{sec:mhdrotor}, with results comparable to those shown in \citet{toth00}. Finally the algorithm was tested on the
Orszag-Tang vortex problem (\S\ref{sec:orstang}) which has
been widely used as a benchmark for MHD codes. The SPMHD results were in good
agreement with those presented elsewhere. This test again highlighted the need for a
slightly larger number of neighbours, in this case to remove spurious effects related to the initial
lattice arrangement and to reduce the magnitude of the divergence errors. The hyperbolic/parabolic divergence cleaning
was found to produce only a small reduction in the divergence errors, again
highlighting the need for some form of sub-timestep cleaning (for example using the
projection method).

 An issue which has not been discussed in this paper, but which needs to be addressed elsewhere, is the tendency of SPH particles merge together at short separations due to the fact that the force tends to zero near the origin of the cubic spline kernel. In particular this problem can become more acute as the number of neighbours is increased (as is required in order to maintain the divergence constraint in MHD). This instability is well known but is not necessarily noticeable in SPH simulations, particularly in 3 dimensions, as it simply leads to a lower effective resolution. Whilst \citet{tc92} propose a simple solution whereby the kernel gradient in the cubic spline is modified slightly whilst retaining the usual kernel for the density evaluation, it is not clear what effect this has on the evolution, particularly when using the variable smoothing length formalism which we have described here and in paper~II. We therefore feel that this problem in particular warrants further attention.

 Finally it is worth commenting on the ability of the algorithm as it stands to treat `real' astrophysical MHD problems. The crucial issue here is the degree to which the divergence constraint can be maintained. Of the methods examined in this paper the most promising is the projection method using the Biot-Savart law since it is the only method which guarantees a zero divergence. Efficient implementation of this method in three dimensions requires use of a tree code (or similar) to solve the resulting Poisson-type equation similar to that used to compute the gravitational force. Note that the treecode implementation differs slightly from the usual gravity tree since the source term of the Poisson equation in this case is a vector quantity. Periodic boundary conditions add a further complication although again methods used for gravity can be easily adapted. Secondly the issue regarding particle merging discussed above needs to be addressed to be able to usefully increase the neighbour number. Thus, whilst many improvements could still be made to the algorithm, the results presented in this paper suggest that the method is ripe for application to problems of current theoretical interest, such as that of star formation.
  
\section*{Acknowledgements} 
DJP would like to especially thank Prof J.E. Pringle and Dr. M.R. Bate for numerous useful discussions. This work has been supported by PPARC, the Commonwealth Scholarship Commission and the Cambridge Commonwealth Trust.
\appendix

\bibliography{sph,mhd}

\label{lastpage}
\enddocument